\pdfoutput=1
\RequirePackage{ifpdf}
\ifpdf 
\documentclass[pdftex]{sigma}
\else
\documentclass{sigma}
\fi

\numberwithin{equation}{section}

\usepackage{braket}
\usepackage{physics}
\usepackage{tikz}
\usetikzlibrary{patterns}

\newcommand{\Hvier}{$H^{4}$}
\newcommand{\Hsechs}{$H^{6}$}
\newcommand{\Z}{\mathbb{Z}}

\newcommand{\N}{\mathbb{N}}

\newcommand{\SymPy}{\texttt{SymPy}}

\DeclareMathOperator{\Mob}{M\ddot{o}b}

\DeclareMathOperator{\Id}{Id}

\newcommand{\QV}{Q_{\rm V}}

\newcommand{\Fp}[1]{F^{(+)}_{#1}}
\newcommand{\Fm}[1]{F^{(-)}_{#1}}

\newcommand{\Fppp}{\Fp{n}\Fp{m}}
\newcommand{\Fpmm}{\Fp{n}\Fm{m}}
\newcommand{\Fmpm}{\Fm{n}\Fp{m}}
\newcommand{\Fmmp}{\Fm{n}\Fm{m}}

\renewcommand{\epsilon}{\varepsilon}
\renewcommand{\imath}{\mathrm{i}}

\makeatletter
\renewcommand{\pdv}[2]{\begingroup
 \@tempswafalse\toks@={}\count@=\z@
 \@for\next:=#2\do
 {\expandafter\check@var\next\@nil
 \advance\count@\der@exp
 \if@tempswa
 \toks@=\expandafter{\the\toks@\,}%
 \else
 \@tempswatrue
 \fi
 \toks@=\expandafter{\the\expandafter\toks@\expandafter\partial\der@var}}%
 \frac{\partial\ifnum\count@=\@ne\else^{\number\count@}\fi#1}{\the\toks@}%
 \endgroup}
\def\check@var{\@ifstar{\mult@var}{\one@var}}
\def\mult@var#1#2\@nil{\def\der@var{#2^{#1}}\def\der@exp{#1}}
\def\one@var#1\@nil{\def\der@var{#1}\chardef\der@exp\@ne}
\makeatother
\newtheorem{Theorem}{Theorem}[section]
 { \theoremstyle{definition}
\newtheorem{Remark}[Theorem]{Remark} }

\begin{document}

\allowdisplaybreaks

\newcommand{\arXivNumber}{1705.00298}

\renewcommand{\thefootnote}{}

\renewcommand{\PaperNumber}{004}

\FirstPageHeading

\ShortArticleName{A Non-Autonomous Approach to the Hietarinta Equation}

\ArticleName{Reconstructing a Lattice Equation:\\ a Non-Autonomous Approach\\ to the Hietarinta Equation\footnote{This paper is a~contribution to the Special Issue on Symmetries and Integrability of Dif\/ference Equations. The full collection is available at \href{http://www.emis.de/journals/SIGMA/SIDE12.html}{http://www.emis.de/journals/SIGMA/SIDE12.html}}}

\Author{Giorgio GUBBIOTTI~$^{\dag\ddag}$ and Christian SCIMITERNA~$^{\ddag}$}

\AuthorNameForHeading{G.~Gubbiotti and C.~Scimiterna}

\Address{$^{\dag}$~School of Mathematics and Statistics, F07, The University of Sydney, \\
\hphantom{$^{\dag}$}~New South Wales 2006, Australia}
\EmailD{\href{mailto:giorgio.gubbiotti@sydney.edu.au}{giorgio.gubbiotti@sydney.edu.au}}

\Address{$^{\ddag}$~Dipartimento di Matematica e Fisica, Universit\`a degli Studi Roma Tre\\
\hphantom{$^{\ddag}$}~and Sezione INFN di Roma Tre, Via della Vasca Navale 84, 00146 Roma, Italy}
\EmailD{\href{mailto:gubbiotti@mat.uniroma3.it}{gubbiotti@mat.uniroma3.it}, \href{mailto:scimiterna@fis.uniroma3.it}{scimiterna@fis.uniroma3.it}}

\ArticleDates{Received April 30, 2017, in f\/inal form December 15, 2017; Published online January 09, 2018}

\Abstract{In this paper we construct a non-autonomous version of the Hietarinta equation [Hietarinta~J., \textit{J.~Phys.~A: Math. Gen.} \textbf{37} (2004), L67--L73] and study its integrability properties. We show that this equation possess linear growth of the degrees of iterates, generalized symmetries depending on arbitrary functions, and that it is Darboux integrable. We use the f\/irst integrals to provide a~general solution of this equation. In particular we show that this equation is a~sub-case of the non-autonomous $Q_{\rm V}$ equation, and we provide a non-autonomous M\"obius transformation to another equation found in [Hietarinta~J., \textit{J.~Nonlinear Math. Phys.} \textbf{12} (2005), suppl.~2, 223--230] and appearing also in Boll's classif\/ication [Boll~R., Ph.D.~Thesis, Technische Universit\"at Berlin, 2012].}


\Classification{37K10; 37K35; 37L20; 37L60; 39A14; 39A22}

\renewcommand{\thefootnote}{\arabic{footnote}}
\setcounter{footnote}{0}

\section{Introduction}

\looseness=-1 Since its introduction the consistency around the cube (CAC) has been a source of many results in the classif\/ication of nonlinear integrable partial dif\/ference equations on a quad graph. The importance of this criterion relies on the fact that it ensures the existence of B\"acklund transformations~\cite{BobenkoSuris2002,Bridgman2013,DoliwaSantini1997,Nijhoff2002,Nijhoff2001} and, as a consequence, of Lax pairs. However \cite{Yamilov2006} Lax pairs and B\"acklund transforms are associated with both linearizable and integrable equations. Let us point out that to be \emph{bona fide} a Lax pair has to give rise to a genuine spectral problem \cite{CalogeroDeGasperisIST_I}, otherwise the Lax pair is a \emph{fake Lax pair} \cite{HayButler2013,HayButler2015, CalogeroNucci1991,Hay2009,Hay2011}. A fake Lax pair is useless in proving (or dis\-pro\-ving) the integrability, since it can be equally found for integrable and non-integrable equations. In the linearizable case Lax pairs are fake, even {though proving this it is usually nontrivial}~\cite{GSL_Gallipoli15}.

\begin{figure}[bthp] \centering
 \begin{tikzpicture}
 \node (x1) at (0,0) [circle,fill,label=-135:$x$] {};
 \node (x4) at (0,2.5) [circle,fill,label=135:$x_{1}$] {};
 \node (x2) at (2.5,0) [circle,fill,label=-45:$x_{2}$] {};
 \node (x3) at (2.5,2.5) [circle,fill,label=45:$x_{12}$] {};
 \draw [thick] (x2) to node[below] {$\alpha_{1}$} (x1);
 \draw [thick] (x4) to node[above] {$\alpha_{1}$} (x3);
 \draw [thick] (x3) to node[right] {$\alpha_{2}$} (x2);
 \draw [thick] (x1) to node[left] {$\alpha_{2}$} (x4);
 \end{tikzpicture}
 \caption{The purely geometric quad graph not embedded in any lattice.}\label{fig:geomquad}
\end{figure}
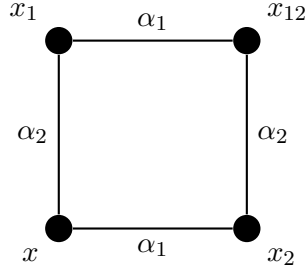

The f\/irst attempt to classify all the multi-af\/f\/ine partial dif\/ference equations def\/ined on the quad graph and possessing CAC was carried out
in~\cite{ABS2003}. There the equation on the quad graph was treated as a geometric object not embedded in any $\Z^{2}$-lattice, as displayed in Fig.~\ref{fig:geomquad}. The quad-equation is an expression of the form
\begin{gather}\label{eq:quadequa}
 Q( x,x_{1},x_{2},x_{12};\alpha_{1},\alpha_{2}) =0,
\end{gather}
connecting some \emph{a priori} independent f\/ields $x$, $x_{1}$, $x_{2}$, $x_{12}$ assigned to the vertices of the quad graph, see Fig.~\ref{fig:geomquad}.
$Q$ is assumed to be a \emph{multi-affine} polynomial in $x$, $x_{1}$, $x_{2}$, $x_{12}$ and, as shown in Fig.~\ref{fig:geomquad}, $\alpha_{1}$ and $\alpha_{2}$ are parameters assigned to the edges of the quad graph.

\begin{figure}[htbp]
 \centering
 \begin{tikzpicture}[auto,scale=0.8]
 \node (x) at (0,0) [circle,fill,label=-135:$x$] {};
 \node (x1) at (4,0) [circle,fill,label=-45:$x_{1}$] {};
 \node (x2) at (1.5,1.5) [circle,fill,label=-45:$x_{2}$] {};
 \node (x3) at (0,4) [circle,fill,label=135:$x_{3}$] {};
 \node (x12) at (5.5,1.5) [circle,fill,label=-45:$x_{12}$] {};
 \node (x13) at (4,4) [circle,fill,label=-45:$x_{13}$] {};
 \node (x23) at (1.5,5.5) [circle,fill,label=135:$x_{23}$] {};
 \node (x123) at (5.5,5.5) [circle,fill,label=45:$x_{123}$] {};
 \node (A) at (2.75,0.75) {$A$};
 \node (Aq) at (2.75,4.75) {$\bar A$};
 \node (B) at (0.75,2.75) {$B$};
 \node (Bq) at (4.75,2.75) {$\bar B$};
 \node (C) at (2,2) {$C$};
 \node (Cq) at (3.5,3.5) {$\bar C$};
 \draw (x) -- node[below]{$\alpha_{1}$} (x1)
 -- node[right] {$\alpha_{2}$} (x12) -- node[right] {$\alpha_{3}$} (x123)
 -- node[above] {$\alpha_{1}$} (x23) -- node[left] {$\alpha_{2}$} (x3) -- node[left] {$\alpha_{3}$} (x);
 \draw (x3) -- node[above]{$\alpha_{1}$} (x13) -- node[right]{$\alpha_{3}$} (x1);
 \draw (x13) -- node[right]{$\alpha_{2}$} (x123);
 \draw [dashed] (x) --node[left]{$\alpha_{2}$} (x2)
 -- node[above] {$\alpha_{1}$} (x12);
 \draw [dashed] (x2) --node[left] {$\alpha_{3}$} (x23);
 \draw [dotted,thick] (A) to (Aq);
 \draw [dotted,thick] (B) to (Bq);
 \draw [dotted,thick] (C) to (Cq);
 \end{tikzpicture}
 \caption{Equations on a cube.} \label{fig:cube2}
\end{figure}
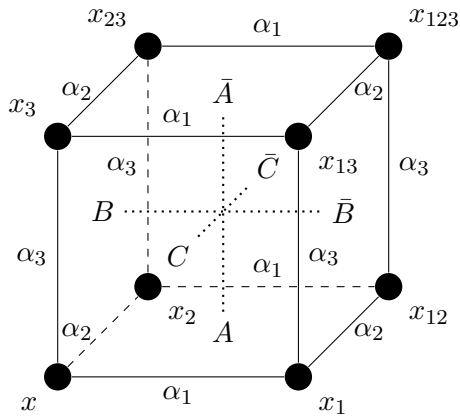

In this setting, we def\/ine the consistency around the cube as follows: assume we are given six quad-equations
\begin{subequations} \label{eq:system}
 \begin{gather}
 A(x,x_{1},x_{2},x_{12};\alpha_1, \alpha_2) =0, \label{eq:Aeq} \\
 \bar{A}(x_{3},x_{13},x_{23},x_{123};\alpha_1, \alpha_2) =0, \label{eq:Abeq} \\
 B(x,x_{2},x_{3},x_{23};\alpha_2, \alpha_3) =0, \label{eq:Beq} \\
 \bar{B}(x_{1},x_{12},x_{13},x_{123};\alpha_2, \alpha_3)=0, \label{eq:Bbeq} \\
 C(x,x_{1},x_{3},x_{13};\alpha_1, \alpha_3) =0, \label{eq:Ceq} \\
 \bar{C}(x_{2},x_{12},x_{23},x_{123};\alpha_1, \alpha_3) =0, \label{eq:Cbeq}
 \end{gather}
\end{subequations}
arranged on the faces of a cube as in Fig.~\ref{fig:cube2}. Using the system \eqref{eq:system} we can compute~$x_{12}$,~$x_{23}$ and~$x_{13}$ from \eqref{eq:Aeq}, \eqref{eq:Beq} and \eqref{eq:Ceq} respectively. Then substituting these values into~\eqref{eq:Abeq}, \eqref{eq:Bbeq} and \eqref{eq:Cbeq} we have three dif\/ferent ways to compute $x_{123}$. If these three dif\/ferent ways of computing $x_{123}$ agree we say that the system \eqref{eq:system} possesses the \emph{consistency around the cube}.

In \cite{ABS2003} the classif\/ication was carried out up to the action of a general M\"obius transformation and up to point transformations of the edge parameters, with the additional assumptions:
\begin{enumerate}\itemsep=0pt
 \item All the faces of the cube in Fig.~\ref{fig:cube2} carry the same equation up to the edge parameters.
 \item The quad-equation \eqref{eq:quadequa} possesses the $D_{4}$ discrete symmetries{\samepage
 \begin{gather}\label{eq:squaresymm}
 Q( x,x_{1},x_{2},x_{12};\alpha_{1},\alpha_{2} )= \mu Q( x,x_{2},x_{1},x_{12};\alpha_{2},\alpha_{1} )= \mu'
 Q( x_{1},x,x_{12},x_{2};\alpha_{1},\alpha_{2}),\!\!\!
 \end{gather}
 where $\mu,\mu'\in\Set{\pm1}$.}

 \item The system \eqref{eq:system} possesses the \emph{tetrahedron property}, i.e., $x_{123}$ is independent of~$x$
 \begin{gather*}
 x_{123} = x_{123}( x,x_{1},x_{2},x_{3};\alpha_{1},\alpha_{2},\alpha_{3} ) \implies \pdv{x_{123}}{x} = 0. 
 \end{gather*}
\end{enumerate}
The results were three classes of discrete autonomous equations with these properties: the $H$ equations, $Q$ equations and the $A$ equations. However the $A$ equations can be transformed in particular cases of the $Q$ equations through non-autonomous M\"obius transformation. Therefore the $A$ equations are usually removed from the general classif\/ication.

After the introduction of the ABS equations J.~Hietarinta tried to weaken the hypotheses of this classif\/ication. First in~\cite{Hietarinta2004} he made a new search for new equations with no assumption about the symmetry and the tetrahedron property. Therein he obtained the following new equation
\begin{gather}
 \frac{x+e_{2}}{x+e_{1}}\frac{x_{12}+o_{2}}{x_{12}+o_{1}} = \frac{x_{1}+e_{2}}{x_{1}+o_{1}}\frac{x_{2}+o_{2}}{x_{2}+e_{1}}, \label{eq:j1}
\end{gather}
where $e_{i}$ and $o_{i}$ are constants. We will refer to this equation as the \emph{Hietarinta equation}. It was later proved that the Hietarinta
equation~\eqref{eq:j1} embedded into a $\Z^{2}$-lattice with the standard embedding
\begin{gather}
 x \to u_{n,m}, \qquad x_{1} \to u_{n+1,m}, \qquad x_{2} \to u_{n,m+1}, \qquad x_{12} \to u_{n+1,m+1}, \label{eq:trivialemb}
\end{gather}
is \emph{linearizable} \cite{Ramanietal2006}. In a subsequent paper \cite{Hietarinta2005} J.~Hietarinta made a new classif\/ication adding the $D_{4}$ discrete symmetries~\eqref{eq:squaresymm} and he found three ``new'' equations, all linearizable.

Releasing the hypothesis that every face of the cube carried
the same equation, in \cite{ABS2009} were presented some
new equations without classif\/ication purposes.
A complete classif\/ication in this extended setting
was then accomplished by Boll in a series of papers~\cite{Boll2011,Boll2012a},
culminating in his Ph.D.~Thesis~\cite{Boll2012b}.
In these papers the classif\/ication of all
the consistent sextupletts of partial dif\/ference equations
on the quad graph, i.e., systems of the form \eqref{eq:system}, has been carried out. The only technical assumption used in by Boll is the tetrahedron property. The obtained equations may fall into three disjoint families depending on their bi-quadratics
\begin{gather}
 h_{ij}=\pdv{Q}{y_{k}}\pdv{Q}{y_{l}}-Q\pdv{Q}{y_{k},y_{l}}, \qquad 		Q=Q ( y_{1},y_{2},y_{3},y_{4};\alpha_1,\alpha_2 ), \label{eq:biquadr}
\end{gather}
where we use a special notation for variables of $Q$, and the pair $\{k,l\}$ is the complement of the pair $\{i,j\}$
in $ \{ 1,2,3,4 \}$.
A bi-quadratic is called
\emph{degenerate} if it contains linear factors of the form $y_{i}-c$,
where $c$ is a constant, otherwise a bi-quadratic
is called \emph{non-degenerate}.
The three families are characterized by how many bi-quadratics
are degenerate:
\begin{itemize}\itemsep=0pt
 \item $Q$-type equations: all the bi-quadratics are non-degenerate,
 \item $H^{4}$-type equations: four bi-quadratics are degenerate,
 \item $H^{6}$-type equations: all of the six bi-quadratics are degenerate.
\end{itemize}
Let us notice that the $Q$ family is the same as the one introduced in~\cite{ABS2003}.
The $H^{4}$ equations are divided into two subclasses: \emph{rhombic}
and \emph{trapezoidal}, depending on their discrete symmetries.

\begin{figure}[htpb]
\centering
\begin{tikzpicture}[scale=2.5]
 \draw [pattern=north west lines,thick] (0,0) rectangle (1,1);
 \draw [pattern=north east lines,thick] (1,1) rectangle (2,2);
 \draw [pattern=horizontal lines,thick] (1,0) rectangle (2,1);
 \draw [pattern=vertical lines,thick] (0,1) rectangle (1,2);
 \foreach \x in {0,...,2}{
 \foreach \y in {0,...,2}{
 \node[draw,circle,inner sep=2pt,fill] at (\x,\y) {};
 }
 }
 \node[below left] at (0,0) {$x$};
 \node[below] at (1,0) {$x_1$};
 \node[below right] at (2,0) {$x$};
 \node[left] at (0,1) {$x_2$};
 \node[below left] at (1,1) {$x_{12}$};
 \node[right] at (2,1) {$x_2$};
 \node[above left] at (0,2) {$x$};
 \node[above] at (1,2) {$x_1$};
 \node[above right] at (2,2) {$x$};
 \node[] at (1/2,1/2) {$Q$};
 \node[] at (1/2+1,1/2) {$|Q$};
 \node[] at (1/2,1/2+1) {$\underline{Q}$};
 \node[] at (1/2+1,1/2+1) {$|\underline{Q}$};
\end{tikzpicture}
\caption{The ``four stripe'' lattice.}\label{fig:elcell}
\end{figure}
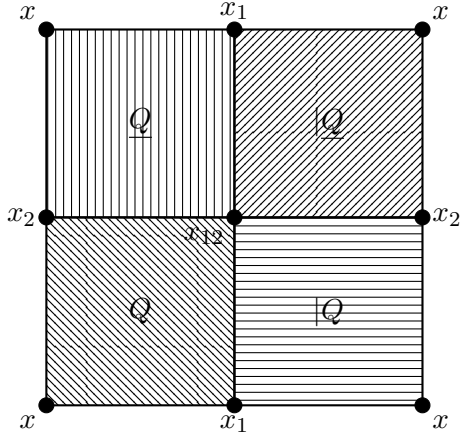

We remark that the classif\/ication results
of \cite{Boll2011,Boll2012a,Boll2012b} hold locally
in the sense that they relate to a single quadrilateral
cell or a single cube displayed in Figs.~\ref{fig:geomquad} and \ref{fig:cube2}.
The important problem of embedding these results into
a two- or three-dimensional lattice, with preservation of the three-dimensional consistency
condition, was discussed in \cite{ABS2009,Xenitidis2009}
by using the concept of a Black and White lattice.
One way to solve this problem is
to embed \eqref{eq:quadequa} into a $\Z^2$-lattice with an elementary cell of size greater
than one. In this case, the quad-equation
\eqref{eq:quadequa} can be extended to a lattice,
and the lattice equation becomes integrable or
linearizable. To this end, following
\cite{Boll2011,Boll2012a,Boll2012b},
we ref\/lect the square with respect to the normal
to its right and top sides and then complete a
$2\times2$ lattice by again ref\/lecting one of the
obtained squares in the other direction.
Such procedure is graphically described in Fig.~\ref{fig:elcell}.

It corresponds to constructing three equations obtained from \eqref{eq:quadequa} by f\/lipping its arguments
 \begin{gather*}
 Q =Q(x,x_{1},x_{2},x_{12};\alpha_{1},\alpha_{2}) =0, \\
 |Q =Q(x_{1},x,x_{12},x_{2};\alpha_{1},\alpha_{2}) =0, \\
 \underline{Q} = Q(x_{2},x_{12},x,x_{1};\alpha_{1},\alpha_{2}) =0, \\
 |\underline{Q} = Q(x_{12},x_{2},x_{1},x;\alpha_{1},\alpha_{2}) =0.
 \end{gather*}
By paving the whole $\Z^{2}$ with such equations, we
get a partial dif\/ference equation which can be in
principle studied using known methods.
Since \emph{a priori}
$Q\neq |Q \neq \underline{Q} \neq |\underline{Q}$,
the obtained lattice will be a four stripe lattice, i.e., an extension of the Black and White
lattice considered in~\cite{ABS2009,HietarintaViallet2012,Xenitidis2009}. This gives rise to lattice equations with two-periodic
coef\/f\/icients for an unknown function $u_{n,m}$, with $(n,m)\in\Z^{2}$
\begin{gather}
\Fppp Q(u_{n,m},u_{n+1,m},u_{n,m+1},u_{n+1,m+1};\alpha_{1},\alpha_{2})\nonumber \\
\qquad{}+\Fmpm|Q (u_{n,m},u_{n+1,m},u_{n,m+1},u_{n+1,m+1};\alpha_{1},\alpha_{2})\nonumber \\
\qquad{}+\Fpmm\underline{Q} (u_{n,m},u_{n+1,m},u_{n,m+1},u_{n+1,m+1};\alpha_{1},\alpha_{2})\nonumber \\
\qquad{}+\Fmmp|\underline{Q} (u_{n,m},u_{n+1,m},u_{n,m+1},u_{n+1,m+1};\alpha_{1},\alpha_{2}) =0, \label{eq:dysys3}
\end{gather}
where
\begin{gather}
 F_{k}^{(\pm)} = \frac{1\pm\left( -1 \right)^{k}}{2}. \label{eq:fk}
\end{gather}
This explicit formula was f\/irst presented in \cite{GSL_Gallipoli15}.
We f\/inally remark that the above construction can be
carried out also at the level of the consistency cube displayed
in Fig.~\ref{fig:cube2} which is then embedded in a three dimensional
lattice $\Z^{3}$ with coordinates $ ( n,m,p )$.
The outcome of the procedure is a sextuplet in which
$\bar{A}=T_{p}A$, $\bar{B}=T_{n}B$ and $\bar{C}=T_{m}C$
where $T_n h_{n,m,p}=h_{n+1,m,p}$, $T_m h_{n,m,p}=h_{n,m+1.p}$ and
$T_{p} h_{n,m,p}=h_{n,m,p+1}$ are translation operators.
Therefore on the lattice the sextuplet becomes, as a matter of fact,
a triplet of equations.
For this reason when dealing with consistent equations embedded
on the lattice we will speak about a triplet of equations.
For more details on the construction of equations on the lattice
from the single cell equations,
we refer to the original papers \cite{Boll2011,Boll2012a,Boll2012b, Xenitidis2009},
to the Appendix in~\cite{GSL_general} and to~\cite{GubbiottiPhD2017}.

We remark that the construction outlined above can be applied
to every consistent system of quad-equations~\eqref{eq:system} when
one of the equations does not possess the $D_{4}$ discrete symmet\-ries~\eqref{eq:squaresymm}.
Given an equation possessing the $D_{4}$ discrete symmetries~\eqref{eq:squaresymm}, it can be shown that it reduces to standard
embedding~\eqref{eq:trivialemb}, see, e.g.,~\cite{GubbiottiPhD2017, GSL_general}.

A detailed study of all the lattice equations derived
from the \emph{rhombic} $H^{4}$
family, including the construction of their three-leg forms,
Lax pairs, B\"acklund transformations and inf\/inite hierarchies
of generalized symmetries,
has been presented in~\cite{Xenitidis2009}.
There are plenty of results about
the~$Q$ and the rhombic~$H^{4}$ equations.
On the contrary,
besides the CAC property
little is known about the integrability features
of the \emph{trapezoidal} $H^{4}$ equations
and of the~$H^{6}$ equations.
These equations where thoroughly studied
in a series of papers
\cite{GSL_general,GSL_Gallipoli15,GSL_Pavel,GSL_symmetries,GSL_QV,GSY_DarbouxII,GSY_DarbouxI}
with some unexpected results.
In~\cite{GSL_general} the algebraic entropy
\cite{BellonViallet1999,HietarintaViallet2007,Viallet2006,Viallet2009}
of the trapezoidal \Hvier\ and the \Hsechs\ equations was computed.
The result of this computation showed that the \emph{rate of growth}
of all the trapezoidal~$H^{4}$
and of all~$H^{6}$ equations is \emph{linear}.
According to the \emph{algebraic entropy conjecture}~\cite{FalquiViallet1993,HietarintaViallet2007}
this fact implies the linearizability.
In~\cite{GSY_DarbouxI}, following the suggestions obtained in~\cite{GSL_Gallipoli15,GSL_Pavel},
it was showed that the trapezoidal~$H^{4}$ equations and \emph{all} the~$H^{6}$ equations are
\emph{Darboux integrable}~\cite{AdlerStartsev1999}.
Finally in~\cite{GSY_DarbouxII} it was shown,
applying a modif\/ication of the procedure presented
in \cite{GarifullinYamilov2012},
that Darboux integrability provides the general
solutions of these equations.
Moreover in~\cite{GSL_Pavel} it was showed that
the three quad-equations found in~\cite{Hietarinta2005}
were Darboux integrable.

In this paper we study a new non-autonomous
version of the Hietarinta equation~\eqref{eq:j1}
obtained using the prescriptions of~\cite{ABS2009,Boll2011,Boll2012a,Boll2012b,Xenitidis2009}.
We then show that this equation possesses properties
analogous to those of the other three equations
considered in~\cite{Hietarinta2005}.

First to construct this new equation we start from the polynomial version of the Hietarinta equation \eqref{eq:j1}
\begin{gather}
 Q = ({x+e_{2}})
 ({x_{1}+o_{1}})
 ({x_{2}+e_{1}})
 ({x_{12}+o_{2}})
 -
 ({x+e_{1}})
 ({x_{1}+e_{2}})
 ({x_{2}+o_{2}})
 ({x_{12}+o_{1}}).
 \label{eq:j1pol}
\end{gather}
This equation does not possess the $D_{4}$
discrete symmetries \eqref{eq:squaresymm},
then in addition to the standard embedding
\eqref{eq:trivialemb} we can consider the non-autonomous
embedding given by formula~\eqref{eq:dysys3}.
Applying formula \eqref{eq:dysys3} to the
Hietarinta equation in polynomial form~\eqref{eq:j1pol}
we obtain the following non-autonomous lattice equation
\begin{gather}
 \big(u_{n,m}+K^{(1)}_{n,m}\big)
 \big(u_{n+1,m}+K^{(1)}_{n+1,m}\big)
 \big(u_{n,m+1}+K^{(1)}_{n,m+1}\big)
 \big(u_{n+1,m+1}+K^{(1)}_{n+1,m+1}\big)\nonumber \\
\quad{} =
 \big(u_{n,m}+K^{(2)}_{n,m}\big)
 \big(u_{n+1,m}+K^{(2)}_{n+1,m}\big)
 \big(u_{n,m+1}+K^{(2)}_{n,m+1}\big)
 \big(u_{n+1,m+1}+K^{(2)}_{n+1,m+1}\big), \!\!\!\!\label{eq:j1na}
\end{gather}
which is similar to the original equation
\eqref{eq:j1pol}, but with the four constants $e_{i}$,
$o_{i}$ replaced by the following \emph{two} two-periodic
functions of the lattice variables
 \begin{gather*}
 K^{(1)}_{n,m} = \Fppp e_{2}+ \Fmpm o_{1} + \Fpmm e_{1}+ \Fmmp o_{2} , \label{eq:o1} \\
 K^{(2)}_{n,m} = \Fppp e_{1}+ \Fmpm e_{2} +\Fpmm o_{2} + \Fmmp o_{1}. \label{eq:o2}
 \end{gather*}
We will call equation \eqref{eq:j1na} the \emph{non-autonomous
Hietarinta equation}.
We remark that the non-autonomous equation \eqref{eq:j1na} depends
on the same number of parameters as its autonomous counterpart \eqref{eq:j1pol}.
Then it is easy to show that the non-autonomous Hietarinta
equation \eqref{eq:j1na} does not reduce to the autonomous one \eqref{eq:j1pol}
for special values of the parameters.
As a matter of fact equation \eqref{eq:j1na} is not a non-autonomous
extension of the Hietarinta equation, but it is a~dif\/ferent
embedding of the same single-cell equation~\eqref{eq:j1pol}
which produces a non-autonomous dynamical system.

Following the procedure outlined in~\cite{Boll2011,Boll2012a,Boll2012b}
and using the notations of Appendix~A in~\cite{GSL_general} we now explain
how to obtain the consistency around the cube for the
non-autonomous Hietarinta equation \eqref{eq:j1na} and discuss its
properties.
First we denote by
\begin{gather*}
 \widehat{Q}_{n,m} = \widehat{Q}_{n,m} ( u_{n,m},u_{n+1,m},u_{n,m+1},u_{n+1,m+1};e_{1},o_{1},e_{2},o_{2} ) 
\end{gather*}
the dif\/ference between the left- and the right-hand side in~\eqref{eq:j1na}. As explained above we add a~third direction $p$ to the f\/ield $u_{n,m}$
\begin{gather*}
 u_{n,m} \to u_{n,m,p}. 
\end{gather*}
Then from the sextuplet presented in \cite{Hietarinta2004}
and the construction in \cite{Boll2011,Boll2012a,Boll2012b,GSL_general}
we have the following triplet of equations
\begin{subequations} \label{eq:j1nasystem}
 \begin{gather}
 A= \widehat{Q}_{n,m,p}( u_{n,m,p},u_{n+1,m,p},u_{n,m+1,p},u_{n+1,m+1,p};e_{1},o_{1},e_{2},o_{2} ), \label{eq:j1Aeq} \\
 B= \widehat{Q}_{n,m,p}( u_{n,m,p},u_{n,m+1,p},u_{n,m,p+1},u_{n,m+1,p+1};e_{2},o_{2},e_{3},o_{3} ), \label{eq:j1Beq} \\
 C= \widehat{Q}_{n,m,p}( u_{n,m,p},u_{n+1,m,p},u_{n,m,p+1},u_{n+1,m,p+1};e_{1},o_{1},e_{3},o_{3} ), \label{eq:j1Ceq}
 \end{gather}
\end{subequations}
being $\bar{A}=T_{p}A$, $\bar{B}=T_{n}B$ and $\bar{C}=T_{m}C$
as remarked above.
We emphasize that the explicit dependence on the $p$
variable is introduced in~\eqref{eq:j1nasystem} though the
two-periodic embedding in~$\Z^{3}$~\smash{\cite{Boll2012b,GSL_general}}.
In practice this corresponds to the replace $F^{(\pm)}_{n}F^{(\pm)}_{m}$
with $F^{(\pm)}_{n}F^{(\pm)}_{m}F^{(\pm)}_{p}$ co\-he\-rent\-ly with the
embedding. Consistency can be checked directly: the value of $u_{n+1,m+1,p+1}$ is given by
\begin{gather}
 u_{n+1,m+1,p+1} =F^{(+)}_{n}F^{(+)}_{m}F^{(+)}_{p}\frac{N_{[123]}}{D_{[123]}}
 +F^{(-)}_{n}F^{(+)}_{m}F^{(+)}_{p}\left.\frac{N_{[123]}}{D_{[123]}}\right|_{e_1 \leftrightarrow o_1}
 \nonumber\\
\hphantom{u_{n+1,m+1,p+1} =}{} +F^{(+)}_{n}F^{(-)}_{m}F^{(+)}_{p}\left.\frac{N_{[123]}}{D_{[123]}}\right|_{e_2 \leftrightarrow o_2}
 +F^{(+)}_{n}F^{(+)}_{m}F^{(-)}_{p}\left.\frac{N_{[123]}}{D_{[123]}}\right|_{e_3 \leftrightarrow o_3}
 \nonumber\\
\hphantom{u_{n+1,m+1,p+1} =}{}+F^{(-)}_{n}F^{(-)}_{m}F^{(+)}_{p}\left.\frac{N_{[123]}}{D_{[123]}}
 \right|_{\substack{e_1 \leftrightarrow o_1\\e_2 \leftrightarrow o_2}}
 +F^{(-)}_{n}F^{(+)}_{m}F^{(-)}_{p}\left.\frac{N_{[123]}}{D_{[123]}}
 \right|_{\substack{e_1 \leftrightarrow o_1\\e_3 \leftrightarrow o_3}}
 \nonumber\\
\hphantom{u_{n+1,m+1,p+1} =}{}+F^{(+)}_{n}F^{(-)}_{m}F^{(-)}_{p}\left.\frac{N_{[123]}}{D_{[123]}}
 \right|_{\substack{e_2 \leftrightarrow o_2\\e_3 \leftrightarrow o_3}}
 +F^{(-)}_{n}F^{(-)}_{m}F^{(-)}_{p}\left.\frac{N_{[123]}}{D_{[123]}}
 \right|_{\substack{e_1 \leftrightarrow o_1\\e_2 \leftrightarrow o_2\\e_3 \leftrightarrow o_3}}, \label{eq:j1naunpmppp}
\end{gather}
where
 \begin{gather*}
 N_{[123]}= N_{[123]}( u_{n,m,p},u_{n+1,m,p},u_{n,m+1,p},u_{n,m,p+1};e_{1},e_{2},e_{3},o_{1},o_{2},o_{3} ), \label{eq:N123dep} \\
 D_{[123]}= D_{[123]}( u_{n,m,p},u_{n+1,m,p},u_{n,m+1,p},u_{n,m,p+1};e_{1},e_{2},e_{3},o_{1},o_{2},o_{3} ), \label{eq:D123dep}
 \end{gather*}
are given as in \cite{Hietarinta2004} by
 \begin{gather*}
 N_{[123]}=-u_{n,m,p} (u_{n+1,m,p}+o_1) (u_{n,m+1,p}+o_2) (u_{n,m,p+1}+o_3)(o_1-o_2) (o_2-o_3) (o_3-o_1)
 \nonumber\\
\hphantom{N_{[123]}=}{} +(u_{n+1,m,p}+o_1) (u_{n,m+1,p}+o_2) (u_{n,m,p+1}+o_3)
 \nonumber\\
\hphantom{N_{[123]}=}{} \times
 [ (e_1 e_2+e_3 o_3) o_3 (o_1-o_2)+(e_2 e_3+e_1 o_1) o_1 (o_2-o_3)+(e_3 e_1+e_2 o_2) o_2 (o_3-o_1)]
 \nonumber\\
\hphantom{N_{[123]}=}{} +(u_{n,m,p}+e_3) (u_{n+1,m,p}+o_1) (u_{n,m+1,p}+o_2)
 o_3 (o_1-o_2) (e_2-o_3) (o_3-e_1)
 \nonumber\\
\hphantom{N_{[123]}=}{} +(u_{n,m,p}+e_1) (u_{n,m+1,p}+o_2) (u_{n,m,p+1}+o_3)o_1 (o_2-o_3) (e_3-o_1) (o_1-e_2)
 \nonumber\\
\hphantom{N_{[123]}=}{} +(u_{n,m,p}+e_2) (u_{n,m,p+1}+o_3) (u_{n+1,m,p}+o_1) o_2 (o_3-o_1) (e_1-o_2) (o_2-e_3),\!\!\!\! \label{eq:N123} \\
 D_{[123]}
 =(u_{n+1,m,p}+o_1) (u_{n,m+1,p}+o_2) (u_{n,m,p+1}+o_3)
 \nonumber\\
\hphantom{D_{[123]} =}{} \times [ (e_1 e_2+e_3 o_3) (o_2-o_1) + (e_2 e_3+e_1 o_1) (o_3-o_2)+(e_3 e_1+e_2 o_2) (o_1-o_3)]
 \nonumber\\
\hphantom{D_{[123]} =}{} + (u_{n,m,p}+e_3)(u_{n+1,m,p}+o_1) (u_{n,m+1,p}+o_2)(o_1-o_2) (e_1-o_3) (e_2-o_3)
 \nonumber\\
\hphantom{D_{[123]} =}{} + (u_{n,m,p}+e_1)(u_{n,m+1,p}+o_2) (u_{n,m,p+1}+o_3) (o_2-o_3) (e_2-o_1) (e_3-o_1)
 \nonumber\\
\hphantom{D_{[123]} =}{} + (u_{n,m,p}+e_2)(u_{n,m,p+1}+o_3) (u_{n+1,m,p}+o_1) (o_3-o_1) (e_3-o_2) (e_1-o_2), \label{eq:D123}
 \end{gather*}
and the notation $A|_{p\leftrightarrow q}$ means that the two parameters must be exchanged. It is clear that $u_{n+1,m+1,p+1}$ as given by~\eqref{eq:j1naunpmppp} depends explicitly on $u_{n,m,p}$, so that the triplet~\eqref{eq:j1nasystem} does not posses the tetrahedron property.

For the rest of this paper we will make a
complete study of the non-autonomous
Hietarinta equation \eqref{eq:j1na}.
In Section \ref{sec:entropy} we will show that equation~\eqref{eq:j1na}
possesses linear growth in all directions.
In Section \ref{sec:qvmobius} we identify equation \eqref{eq:j1na}
with a sub-case of the non-autonomous $\QV$ equation
\cite{GSL_QV}. Then we show that the equation \eqref{eq:j1na} can be
reduced to the quad-equation
\begin{gather}
 v_{n,m} v_{n+1,m+1} + v_{n+1,m}v_{n,m+1}=0.
 \label{eq:j2v}
\end{gather}
which was discussed in \cite{Hietarinta2004,Hietarinta2005}.
This f\/inding immediately explains the linearizability
property of this equation and it is \emph{the main result
of this paper}.
Indeed equation \eqref{eq:j2v} is linked to the discrete
wave equation
\begin{gather}
 w_{n+1,m+1}-w_{n+1,m}-w_{n,m+1}+w_{n,m}=0
 \label{eq:dwave}
\end{gather}
through the non-autonomous, transcendental transformation
\begin{gather}
 v_{n,m} = \big[ F^{(+)}_{m}+F^{(-)}_{m}\big(F^{(+)}_{n}-F^{(-)}_{n} \big)\big] e^{w_{n,m}}. \label{eq:transcwave}
\end{gather}
This result is of course quite unexpected.
Then we brief\/ly discuss the r\^ole of the tetrahedron property
in these investigations.
In the subsequent Sections we treat the equation~\eqref{eq:j1na} into detail in order to show all the interesting
properties that were found for the trapezoidal~$H^{4}$ and~$H^{6}$ equations
\cite{GSL_general,GSL_Gallipoli15,GSL_symmetries,GSL_QV,GSY_DarbouxI,GSY_DarbouxII}
with this simple example.
In Section~\ref{sec:darb} we will prove that equation~\eqref{eq:j1na}
is Darboux integrable with f\/irst order f\/irst integrals.
In Section~\ref{sec:gensymm} we will present
the generalized symmetries of the non-autonomous
Hietarinta equation~\eqref{eq:j1na} of every
order by using the f\/irst integrals and the so
called \emph{symmetry drivers}~\cite{Startsev2016}.
In Section~\ref{sec:sol} we use the f\/irst integrals to derive
the general solution of equation~\eqref{eq:j1na}.
In Section~\ref{sec:conclusions} we give some
conclusion and we give an outlook on presented results.

The main objective of this paper is to show how
the Hietarinta equation~\eqref{eq:j1},
behaves under the Boll's embedding,
becoming then the non-autonomous equation~\eqref{eq:j1na}
and then to discuss the property of this new embedding.
In this sense our starting point is similar to those
adopted in \cite{HietarintaViallet2012}, where it
was shown that dif\/ferent consistent embeddings
can produce equations with dif\/ferent properties.

\section{Algebraic entropy}\label{sec:entropy}

\emph{Algebraic entropy} is as a test of integrability for discrete systems. Given a bi-rational map, which can be an ordinary dif\/ference equation, a dif\/ferential dif\/ference equation or even a partial dif\/ference equation, the basic idea is to examine the growth of the degree of its iterates, and extract a canonical quantity, which is an index of complexity of the map. This canonical quantity will be the algebraic entropy (or its avatar the dynamical degree) \cite{BellonViallet1999,Diller1996,FalquiViallet1993,Russakovskii1997,Veselov1992}. In \cite{Tremblay2001,Viallet2006} the method was developed in the case of quad-equations and then used as a classifying tool~\cite{HietarintaViallet2007}. A slight generalization of the method was given in \cite{GSL_general} where it was showed that non-autonomous equations can have more that a single sequence of degrees.

Given a sequence of degrees obtained iterating a rational map
\begin{gather*}
 1,d_{0},d_{1},\dots,d_{k},\dots, 
\end{gather*}
we def\/ine the algebraic entropy of this map to be
\begin{gather}
 \eta = \lim_{k\to\infty}\frac{1}{k}\log d_{k}.
 \label{eq:algentdef}
\end{gather}
This quantity is canonical as it is invariant with respect
to bi-rational transformations.
To give a practical example, the transformation \eqref{eq:transcwave}
which maps equation \eqref{eq:j2v} into the discrete
wave equation \eqref{eq:dwave} does not preserve
\emph{a priori} algebraic entropy, because it is not bi-rational,
see~\cite{Grammaticos2005}.
Geometrically algebraic entropy is deeply
linked with the structure of the singularities
of a discrete system, and in some cases
it can be computed resorting to this structure
\cite{DillerFavre2001,Sakai2001,Takenawa2001,Viallet2015}.

Instead of computing the whole
sequence of iterates which is clearly impossible,
only a f\/inite number of iterates is computed.
Then some tools like generating functions~\cite{Lando2003}
or discrete derivatives are used in order to obtain
the asymptotic behaviour of the series~\cite{GrammaticosHalburdRamaniViallet2009,GubbiottiASIDE16}.

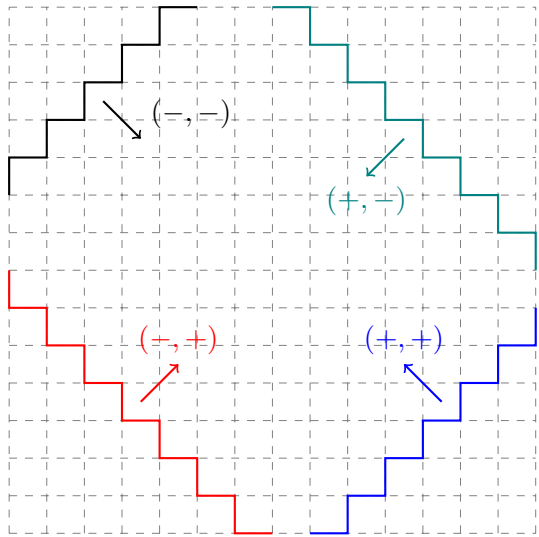
\begin{figure}[hbt]
 \centering
 \begin{tikzpicture}[scale=0.5]
 \draw[style=help lines,dashed] (0,0) grid[step=1cm] (14,14);
 \draw[thick] (0,9)--(0,10)--(1,10)--(1,11)--(2,11)--(2,12)--(3,12)--(3,13)--(4,13)--(4,14)--(5,14);
 \node[above right] at (3+1/2,11-1/2) {$(-,-)$};
 \draw[thick,->] (2+1/2,12-1/2)--(3+1/2,11-1/2);
 \draw[thick,red] (0,7)--(0,6)--(1,6)--(1,5)--(2,5)--(2,4)--(3,4)--(3,3)--(4,3)--(4,2)--(5,2)--(5,1)--(6,1)--(6,0)--(7,0);
 \node[above,red] at (5-1/2,4+1/2) {$(-,+)$};
 \draw[thick,->,red] (4-1/2,3+1/2)--(5-1/2,4+1/2);
 \draw[thick,blue] (8,0)--(9,0)--(9,1)--(10,1)--(10,2)--(11,2)--(11,3)--(12,3)--(12,4)--(13,4)--(13,5)--(14,5)--(14,6);
 \node[above,blue] at (11-1/2,4+1/2) {$(+,+)$};
 \draw[thick,->,blue] (12-1/2,3+1/2)--(11-1/2,4+1/2);
 \draw[thick,teal] (14,7)--(14,8)--(13,8)--(13,9)--(12,9)--(12,10)--(11,10)--(11,11)--(10,11)--(10,12)--(9,12)--(9,13)--(8,13)--(8,14)--(7,14);
 \node[below,teal] at (11-1/2-1,11-1/2-1) {$(+,-)$};
 \draw[thick,->,teal] (11-1/2,11-1/2)--(11-1/2-1,11-1/2-1);
 \end{tikzpicture}
 \caption{Principal growth directions.}
 \label{fig:princgrowth}
\end{figure}

The classif\/ication of lattice equations based on the algebraic
entropy test is:
\begin{enumerate}\itemsep=0pt
 \item[] {\it linear growth}: the equation is linearizable,
 \item[] {\it polynomial growth}: the equation is integrable,
 \item[] {\it exponential growth}: the equation is chaotic.
\end{enumerate}

We have performed the algebraic entropy analysis
in the principal growth directions \cite{Viallet2006},
shown in Fig.~\ref{fig:princgrowth},
of the non-autonomous Hietarinta equation \eqref{eq:j1na}
to identify its behaviour.
To this end we used the \SymPy~\cite{sympy} module
\verb!ae2d.py! \cite{GubbiottiPhD2017, GubHay}.
We found that the non-autonomous Hietarinta equation
\eqref{eq:j1na} is \emph{isotropic}, i.e., its sequence
of degrees is the same in every direction, and
despite being non-autonomous and two-periodic
it possess a single sequence of degrees
given by
\begin{gather}
 1,3,5,7,9,11,13,\dots. \label{eq:oddnumb}
\end{gather}
The sequence \eqref{eq:oddnumb} is asymptotically (in this case exactly) f\/itted by
\begin{gather*}
 d_{k} = 2k +1, \qquad \forall\, k \in\N, 
\end{gather*}
and the algebraic entropy from \eqref{eq:algentdef}
is clearly zero.
Since the growth is linear we expect the non-autonomous Hietarinta
equation \eqref{eq:j1na} to be linearizable.
In the next section we will show why the linearization
arises.

\section{Point transformations and tetrahedron property}\label{sec:qvmobius}

In \cite{GSL_QV} it was proved that there exists a non-autonomous, two-periodic generalization of the $\QV$ equation \cite{Viallet2009} given by
\begin{gather}
 a_{1} u_{{n,m}}u_{{n+1,m}}u_{{n,m+1}}u_{{n+1,m+1}}
 \nonumber\\
 \qquad{} + \big[ { a_{2,0}}- ( -1 ) ^{n}{ a_{2,1}}
 - ( -1 ) ^{m}{a_{2,2}}+ ( -1 ) ^{n+m}{ a_{2,3}}\big] u_{{n,m}}u_{{n,m+1}}u_{{n+1,m+1}}\nonumber \\
\qquad{} + \big[ { a_{2,0}}+( -1 ) ^{n}{ a_{2,1}}- ( -1 ) ^{m}{ a_{2,2}}-
 ( -1 ) ^{n+m}{ a_{2,3}} \big]
 u_{{n+1,m}}u_{{n,m+1}}u_{{n+1,m+1}}\nonumber \\
\qquad{} + \big[ { a_{2,0}}+ ( -1 ) ^{n}{ a_{2,1}}
 + ( -1 ) ^{m}{ a_{2,2}}+ ( -1 ) ^{n+m}{ a_{2,3}} \big]
 u_{{n,m}}u_{{n+1,m}}u_{{n+1,m+1}}\nonumber \\
 \qquad{} + \big[ { a_{2,0}}- ( -1 ) ^{n}{a_{2,1}}+ ( -1 ) ^{m}{ a_{2,2}}
 - ( -1 ) ^{n+m}{ a_{2,3}} \big]
 u_{{n,m}}u_{{n+1,m}}u_{{n,m+1}}\nonumber \\
 \qquad{} + \big[ { a_{3,0}}- ( -1 ) ^{m}{ a_{3,2}} \big] u_{{n,m}}u_{{n+1,m}}
 + \big[ { a_{3,0}}+ ( -1 ) ^{m}{ a_{3,2}} \big] u_{{n,m+1}}u_{{n+1,m+1}}\nonumber \\
 \qquad{} + \big[ { a_{4,0}}- ( -1) ^{n+m}{ a_{4,3}} \big] u_{{n,m}}u_{{n+1,m+1}}
 + \big[ { a_{4,0}}+ ( -1 ) ^{n+m}{ a_{4,3}} \big] u_{{n+1,m}}u_{{n,m+1}}\nonumber \\
 \qquad{} + \big[ { a_{5,0}}- ( -1 ) ^{n}{ a_{5,1}} \big] u_{{n+1,m}}u_{{n+1,m+1}}
 + \big[ { a_{5,0}}+ ( -1 ) ^{n}{ a_{5,1}} \big] u_{{n,m}}u_{{n,m+1}}\nonumber \\
 \qquad{} + \big[ { a_{6,0}}+ ( -1 ) ^{n}{ a_{6,1}}-
 ( -1 ) ^{m}{ a_{6,2}}- ( -1 ) ^{n+m}{ a_{6,3}} \big] u_{{n,m}}\nonumber \\
 \qquad{} + \big[ { a_{6,0}}- ( -1 ) ^{n}{ a_{6,1}}
 - ( -1 ) ^{m}{ a_{6,2}}+ ( -1 ) ^{n+m}{ a_{6,3}} \big] u_{{n+1,m}}\nonumber \\
\qquad{} + \big[ { a_{6,0}}+ ( -1 ) ^{n}{ a_{6,1}}
 + ( -1 ) ^{m}{ a_{6,2}}+ ( -1 ) ^{n+m}{ a_{6,3}} \big] u_{{n,m+1}}\nonumber \\
 \qquad{}+ \big[ { a_{6,0}}- ( -1 ) ^{n}{ a_{6,1}}
 + ( -1 ) ^{m}{ a_{6,2}}- ( -1 ) ^{n+m}{ a_{6,3}}\big] u_{{n+1,m+1}}
 +a_{7}=0. \label{eq:naqv}
\end{gather}
In \cite{GSL_QV} it has been checked heuristically
that this equation is integrable according to the
algebraic entropy test, possessing \emph{quadratic
growth of the degrees of the iterates}, as was later rigorously
shown using the factorization approach in \cite{RobertsTran2017}.
Additionally in~\cite{GSL_QV} it was proved that
equation~\eqref{eq:naqv} \emph{contains as particular cases
all the equations coming from the Boll's classifications},
i.e., the rhombic $H^{4}$ equations, the trapezoidal $H^{4}$
equations, the $H^{6}$ equations and the $Q$ equations.
Upon the substitution
$a_{2,1}=a_{2,2}=a_{2,3}=a_{3,2}=a_{4,3}=a_{5,1}=a_{6,1}=a_{6,2}=a_{6,3}=0$
the non-autonomous $\QV$ equation reduces to
the original $\QV$ equation presented in~\cite{Viallet2009}.
For this reasons equation~\eqref{eq:naqv} has been
called the \emph{non-autonomous, or two-periodic, $\QV$ equation}.

It is just a matter of computation to show that the original
autonomous Hietarinta equation~\eqref{eq:j1} is not
a sub-case of the autonomous~$\QV$ equation, but that
the non-autonomous Hietarinta equation is a sub-case of
the non-autonomous $\QV$ equation \eqref{eq:naqv} with
the following values of the coef\/f\/icients
\begin{subequations} \label{eq:j1cf}
 \begin{gather}
 a_{1} = 0, \\
 a_{2,0} = 0,\qquad
 a_{2,1} = e_{2}-o_{2},\qquad
 a_{2,2} = e_{1}-o_{1},\qquad
 a_{2,3} = e_{1}+o_{1}-o_{2}-e_{2}, \\
 a_{3,0} = -(e_{2}-o_{2}) (e_{1}-o_{1}),\qquad
 a_{3,2} = -(e_{2}+o_{2}) (e_{1}-o_{1}), \\
 a_{4,0} = 0,\qquad
 a_{4,3} = 2 (e_{2} o_{2} - o_{1} e_{1}), \\
 a_{5,0} = (e_{2}-o_{2}) (e_{1}-o_{1}),\qquad
 a_{5,1} = -(e_{2}-o_{2}) (o_{1}+e_{1}), \\
 a_{6,0} = 0,\qquad
 a_{6,1} = -o_{1} e_{1} (e_{2}-o_{2}),\qquad a_{6,2} = -(e_{1}-o_{1}) e_{2} o_{2},\nonumber \\
 a_{6,3} = o_{2} e_{1} o_{1}+e_{2} o_{1} e_{1}-e_{2} o_{2} e_{1}-e_{2} o_{2} o_{1}, \\
 a_{7} = 0.
 \end{gather}
\end{subequations}

In the Introduction we discussed how the non-autonomous
Hietarinta equation \eqref{eq:j1na} comes from the triplet~\eqref{eq:j1nasystem} which does not possess the tetrahedron property.
However equation~\eqref{eq:j1cf} shows that equation~\eqref{eq:j1na}
is a particular case of the non-autonomous $\QV$ equation~\eqref{eq:naqv}
and non-autonomous $\QV$ equation \eqref{eq:naqv} contains
all the equations of Boll's classif\/ication.
This, along with the result of the algebraic entropy obtained in
Section \ref{sec:entropy}, suggests that a transformation mapping
equation \eqref{eq:j1na} into an equation of Boll's classif\/ication might exist.
As stated before all the equations from Boll's classif\/ication come
from sextuplet (triplet on the lattice) possessing the tetrahedron property, therefore the existence
of such transformation will prove that the same equation
can come from a dif\/ferent sextuplet (triplet on the lattice) possessing tetrahedron property.
This turns out to be true since it can be shown that the non-autonomous,
two-periodic M\"obius transformation
\begin{gather}
 u_{n, m} = -\frac{\big[ e_{1} F_{n}^{(+)} - o_{1}F_{n}^{(-)}\big( F_{m}^{(+)}-F_{m}^{(-)} \big) \big]v_{n, m}
 +e_{2}F_{m}^{(+)}+o_{2}F_{m}^{(-)}}{
 \big[ \big( F_{n}^{(+)}-F_{n}^{(-)} \big)F_{m}^{(+)}+F_{m}^{(-)} \big]v_{n, m}+1}
 \label{eq:mobnm}
\end{gather}
maps the non-autonomous Hietarinta equation~\eqref{eq:j1na}
into equation~\eqref{eq:j2v}.
The transformation~\eqref{eq:mobnm} can be interpreted at
level of single cells as a simultaneous transformation of all
the variables associated to the vertices $x$, $x_{1}$, $x_{2}$
and $x_{12}$, i.e., acting with an element of the group
$\Mob^{4}$ \cite{ABS2003,AtkinsonPhD2008,GSL_general}.
Following the explicit formula given in~\cite{GSL_general}
the element of $\Mob^{4}$ corresponding to~\eqref{eq:mobnm}
is given by
\begin{gather}
 x = -\frac{e_{1} X +e_{2}}{X+1},
 \quad
 x_{1} = -\frac{o_{1} X_{1} -e_{2}}{X_{1}-1},
 \quad
 x_{2} = -\frac{e_{1} X_{2}+o_{2}}{X_{2}+1},
 \quad
 x_{12} = -\frac{o_{1} X_{12}+o_{2}}{X_{12}-1}. \label{eq:mob4}
\end{gather}
This brings the Hietarinta equation on the single cell~\eqref{eq:j1pol} into the single cell version of~\eqref{eq:j2v}
\begin{gather*}
 X X_{12} + X_{1}X_{2}=0. 
\end{gather*}

We now discuss what the existence of the transformation
\eqref{eq:mobnm} implies at the level of the consistency around the
cube.
To make the discussion easier to follow, here will work with
the equation on the single cell \eqref{eq:j1pol}.
We will show
that allowing a simultaneous transformation of all the points
$x$, $x_{1}$, $x_{2}$, $x_{3}$, $x_{12}$, $x_{13}$, $x_{23}$
and $x_{123}$, i.e., by using the group $\Mob^{8}$
it is possible to produce a sextuplet of consistent quad-equations
of which \eqref{eq:j2v} is the base equation.
Indeed let us start from the sextuple of equations for the
\eqref{eq:j1} equation given in~\cite{Hietarinta2004}
\begin{subequations} \label{eq:j1hietsystem}
 \begin{gather}
 A = \frac{x+e_{2}}{x+e_{1}}\frac{x_{12}+o_{2}}{x_{12}+o_{1}} -
 \frac{x_{1}+e_{2}}{x_{1}+o_{1}}\frac{x_{2}+o_{2}}{x_{2}+e_{1}}, \label{eq:j1hietA} \\
 \bar{A} = \frac{x_{3}+e_{2}}{x_{3}+e_{13}}\frac{x_{123}+o_{2}}{x_{123}+o_{1}} -
 \frac{x_{13}+e_{2}}{x_{1}+o_{1}}\frac{x_{23}+o_{2}}{x_{23}+e_{1}}, \label{eq:j1hietAt} \\
 B = \frac{x+e_{3}}{x+e_{2}}\frac{x_{23}+o_{3}}{x_{23}+o_{2}} -
 \frac{x_{2}+e_{3}}{x_{2}+o_{2}}\frac{x_{3}+o_{3}}{x_{3}+e_{2}}, \label{eq:j1hietB} \\
 \bar{B} = \frac{x_{1}+e_{3}}{x_{1}+e_{2}}\frac{x_{123}+o_{3}}{x_{123}+o_{2}} -
 \frac{x_{2}+e_{3}}{x_{12}+o_{2}}\frac{x_{13}+o_{3}}{x_{13}+e_{2}}, \label{eq:j1hietBt} \\
 C = \frac{x+e_{3}}{x+e_{1}}\frac{x_{13}+o_{3}}{x_{13}+o_{1}} -
 \frac{x_{1}+e_{3}}{x_{1}+o_{1}}\frac{x_{3}+o_{3}}{x_{3}+e_{1}}, \label{eq:j1hietC} \\
 \bar{C} = \frac{x_{2}+e_{3}}{x_{2}+e_{1}}\frac{x_{123}+o_{3}}{x_{123}+o_{1}} -
 \frac{x_{12}+e_{3}}{x_{12}+o_{1}}\frac{x_{23}+o_{3}}{x_{23}+e_{1}}. \label{eq:j1hietCt}
 \end{gather}
\end{subequations}
Then the $\Mob^{8}$ transformation given by
\begin{gather}
 x = -\frac{e_{1} X +e_{2}}{X+1},
 \qquad
 x_{1} = -\frac{o_{1} X_{1} -e_{2}}{X_{1}-1},
 \qquad
 x_{2} = -\frac{e_{1} X_{2}+o_{2}}{X_{2}+1},
 \qquad
 x_{3} = -\frac{e_{1} X_{3}+e_{2}}{X_3+1},
 \nonumber\\
 x_{12} = -\frac{o_{1} X_{12} +o_{2}}{X_{12}+1},
 \qquad
 x_{13} = -\frac{o_{1} X_{13}-e_{2}}{X_{13}-1},
\qquad
 x_{23} = -\frac{e_{1} X_{23}+o_{2}}{X_{23}+1},
 \nonumber\\
 x_{123} = -\frac{o_{1} X_{123}+o_{2}}{X_{123}+1}. \label{eq:mob8}
\end{gather}
brings the system \eqref{eq:j1hietsystem} into the following one
\begin{subequations} \label{eq:j1hietsystemm}
 \begin{gather}
 A = X X_{12} + X_{1}X_{2}, \label{eq:j1hietAm} \\
 \bar{A} =X_{3} X_{123} + X_{13}X_{23}, \label{eq:j1hietAtm} \\
 B =[ ( { e_{2}} -{ e_{3}})
 ( { e_{1}}-{o_{3}} ) { X_{3}}
 -X ( { e_{2}}-{ o_{3}} ) ( {e_{1}}-{ e_{3}} ) ] { X_{2}}X_{23}\nonumber\\
 \hphantom{B=}{} + ( { e_{3}}-{ o_{2}})
 [ ( { e_{1}}-{ o_{3}} ) {X_{3}}+{e_{2}}-{ o_{3}} ] X X_{23}\nonumber \\
 \hphantom{B=}{}+( {o_{2}}-{ o_{3}} ) ( ( { e_{1}}-{ e_{3}} ) X+{ e_{2}}-{ e_{3}} ) { X_{2}}{ X_{3}}, \label{eq:j1hietBm} \\
 \bar{B} =
 [ ( { o_{1}}-{ o_{3}} )
 ( {e_{2}} -{ e_{3}}) { X_{13}}
 +( { e_{3}}-{ o_{1}} )
 ( { e_{2}}-{ o_{3}} ) { X_{1}} ] { X_{12}}X_{123}\nonumber \\
 \hphantom{\bar{B} =}{} -
 ( { e_{3}}-{ o_{2}} )
 [ ( { o_{1}}-{ o_{3}}) { X_{13}}-{ e_{2}}+{ o_{3}} ]
 X_{1} { X_{123}}\nonumber \\
 \hphantom{\bar{B} =}{}- ( { o_{2}}-{ o_{3}} )
 [ ( { o_{1}} -{ e_{3}}) { X_{1}}-{ e_{2}}+{ e_{3}} ]
 { X_{13}}{ X_{12}}, \label{eq:j1hietBtm} \\
 C =
 e_{3}{ o_{3}}
 [( { X_{3}}+1) { X_{1}} -( X+1 ) { X_{13}}-X+{ X_{3}} ]\nonumber \\
 \hphantom{C=}{} +{ e_{3}}[
 ( X-{ X_{1}} ) { e_{2}}-{ X_{3}}
 ( { X_{1}}+1 ) { e_{1}}+ { o_{1}} ( X+1 ) {X_{13}} ]\nonumber \\
 \hphantom{C=}{}+
 { o_{3}}
 [ (
 { e_{2}}+{ e_{1}} X ) { X_{13}}+{ e_{1}} X-{ e_{2}} { X_{3}}
 - { o_{1}}{ X_{1}} ( { X_{3}}+1 ) ]\nonumber \\
 \hphantom{C=}{}-{o_{1}}( { e_{2}}+{ e_{1}} X ) { X_{13}}
 +e_{1} [ { e_{2}} ({X_{3}}-X )
 +{ o_{1}}{ X_{1}} { X_{3}} ]
 +{ o_{1}} { e_{2}}{ X_{1}}, \label{eq:j1hietCm} \\
 \bar{C} =
 e_{3}
 [ ( { X_{2}}+1 ) { X_{123}}
 -( 1+{X_{23}} ) { X_{12}}
 -{ X_{23}}+{ X_{2}} ] { o_{3}}\nonumber \\
 \hphantom{\bar{C} =}{} +{ e_{3}}[
 ( { o_{2}}+{ e_{1}} {X_{23}} ) { X_{12}}
 -{ o_{1}} ( { X_{2}}+1 ) { X_{123}}
 +{ e_{1}} { X_{23}}-{ o_{2}} { X_{2}} ]\nonumber \\
 \hphantom{\bar{C} =}{}+o_{3}
 [ { o_{1}} ( { X_{23}}+1 ) { X_{12}}
 -( { o_{2}}+{ e_{1}} { X_{2}}) { X_{123}}+
 {o_{2}} { X_{23}}-{ e_{1}} { X_{2}} ]\nonumber \\
 \hphantom{\bar{C} =}{}+{ o_{1}}
 ( { o_{2}}+{ e_{1}} { X_{2}} ) { X_{123}}-{ o_{1}}
 ( { o_{2}}+{ e_{1}} { X_{23}} ) { X_{12}}+{ e_{1}} {
 o_{2}} ( { X_{2}} -{ X_{23}}). \label{eq:j1hietCtm}
 \end{gather}
\end{subequations}
It is easy to show that the system \eqref{eq:j1hietsystemm} do not possess the tetrahedron property. Indeed computing $X_{123}$ we obtain
\begin{gather*}
 X_{123}
 =
 X_{2}\frac{{ o_{2}}-{ o_{3}}}{{ o_{1}}-{ o_{3}}}
 {\frac {\left\{
 \begin{gathered}
 { e_{3}}o_{3}[ ( { X_{1}}-1 ) { X_{3}}+X+{ X_{1}} ]
 \\
 + o_{3}[( { e_{2}}-{ o_{1}} { X_{1}} ) { X_{3}}-{ e_{1}} X-{o_{1}} { X_{1}} ]
 \\
 - e_{3} [ e_{1} ( { X_{1}}-1 )X_{3}
 + e_{2}( X+{ X_{1}} ) ]
 \\
 -{ e_{1}} ( { e_{2}}-{ o_{1}} { X_{1}} ) { X_{3}}
 +{ e_{2}} ( { e_{1}} X+{ o_{1}} { X_{1}} )
 \end{gathered}
 \right\}}{
 \left\{
 \begin{gathered}
 e_{3}o_{3}
 [ (X_{2}-X ) X_{3}-( { X_{2}}+1 )X ]
 \\
 + o_{3}[( o_{2}X-e_{2}{ X_{2}}) { X_{3}}
 +( { o_{2}}+{ e_{1}} { X_{2}} )X ]
 \\
 +e_{3}
 [ { e_{1}} ( X-X_{2} ) { X_{3}}
 +e_{2}( { X_{2}}+1 )X ]
 \\
 -e_{1} (o_{2} X -e_{2} X_{2} ) { X_{3}}
 -e_{2} ( { o_{2}}+{ e_{1}} { X_{2}} ) X
 \end{gathered}\right\}
 }
 }
. 
\end{gather*}

\begin{Remark}
 We remark that the use of $\Mob^{4}$ and $\Mob^{8}$
 transformations like~\eqref{eq:mob4} and~\eqref{eq:mob8}
 is naturally allowed since we are in the framework of the
 non-autonomous CAC construction given in
 \cite{ABS2009,Boll2011,Boll2012a,Boll2012b}.
 Non-autonomous M\"obius transformations can be used also in
 the autonomous case, see \cite{ABS2003,Hietarinta2005}, but
 in non-autonomous case they are the natural group which
 preserves the classif\/ication \cite{GSL_general}.
\end{Remark}

On the other hand equation \eqref{eq:j2v} is a particular case of the $_{1}D_{4}$ equation from \cite{Boll2011,Boll2012a,Boll2012b,GSL_general}
\begin{gather}
\delta_{1}\big(F_{n}^{\left(-\right)}v_{n,m}v_{n,m+1}
 +F_{n}^{\left(+\right)}v_{n+1,m}v_{n+1,m+1}\big) +\delta_{2}\big(F_{m}^{(-)}v_{n,m}v_{n+1,m}+F_{m}^{(+)}v_{n,m+1}v_{n+1,m+1}\big)\nonumber\\
 \qquad{} +v_{n,m}v_{n+1,m+1}+v_{n+1,m}v_{n,m+1}+\delta_{3}=0, \label{eq:1D4}
\end{gather}
with $\delta_{1}=\delta_{2}=\delta_{3}=0$. This means that the non-autonomous Hietarinta equation~\eqref{eq:j1na} is not an independent equation, but is part of Boll's classif\/ication.

In particular this implies that equation~\eqref{eq:j2v} can arise also as bottom equation of a triplet possessing tetrahedron property.
The relevant triplet is that coming from Case~D in Theorem~3.12 in~\cite{Boll2012b} that in our notation reads as
\begin{subequations} \label{eq:j2vsystem2}
 \begin{gather}
 A = v_{n,m,p}v_{n+1,m+1,p}+v_{n+1,m,p}v_{n,m+1,p}, \label{eq:j2vAeq2} \\
 B =\lambda(v_{n,m,p} v_{n,m+1,p}+v_{n,m,p+1} v_{n,m+1,p+1})
 -v_{n,m,p} v_{n,m+1,p+1}-v_{n,m+1,p} v_{n,m,p+1}, \label{eq:j2vBeq2} \\
 C=\frac{1}{\lambda} ( v_{n,m,p} v_{n+1,m,p}+v_{n,m,p+1} v_{n+1,m,p+1} )
 -v_{n,m,p} v_{n+1,m,p+1}-v_{n+1,m,p} v_{n,m,p+1}, \!\!\!\!\label{eq:j2vCeq2}
 \end{gather}
\end{subequations}
Here $\lambda$ is a new non-zero parameter. The triplet \eqref{eq:j2vsystem2} is consistent and possesses the tetrahedron property since
\begin{gather*}
 v_{n+1,m+1,p+1} = -\frac{v_{n+1,m,p} v_{n,m+1,p}}{v_{n,m,p+1}}. 
\end{gather*}

\begin{Remark} We recall that the equation \eqref{eq:j2v} can be derived from another  triplet without the tetrahedron property.  As it was discussed in \cite{Hietarinta2004,Hietarinta2005} the  equation \eqref{eq:j2v} can  thought as $A$ equation of the  following triplet (we are already on the lattice)
 \begin{subequations}\label{eq:j2vsystem}
 \begin{gather}
 A= v_{n,m,p}v_{n+1,m+1,p}+v_{n+1,m,p}v_{n,m+1,p}, \label{eq:j2vAeq} \\
 B= v_{n,m,p}v_{n,m+1,p+1}+v_{n,m+1,p}v_{n,m,p+1}, \label{eq:j2vBeq} \\
 C= v_{n,m,p}v_{n+1,m,p+1}+v_{n+1,m,p}v_{n,m,p+1}, \label{eq:j2vCeq}
 \end{gather}
 \end{subequations}
 which is obtained from \eqref{eq:j2v} properly permuting the shifted variables. The triplet \eqref{eq:j2vsystem} is consistent on the cube
 but with no tetrahedron property since
 \begin{gather*}
 v_{n+1,m+1,p+1} = -\frac{v_{n+1,m,p} v_{n,m+1,p} v_{n,m,p+1}}{v_{n,m}^2}. 
 \end{gather*}
\end{Remark}

This last result is clearly unexpected. \emph{A priori} it would have been very dif\/f\/icult to f\/ind out that the non-autonomous
version of the Hietarinta equation \eqref{eq:j1} was in fact a particular
case of the $_{1}D_{4}$ equation \eqref{eq:1D4},
since the f\/irst one arises from the triplet \eqref{eq:j1nasystem}
without tetrahedron property,
whereas the latter one arises from the triplet \eqref{eq:j2vsystem}
with tetrahedron property.
However, in general, a quad-equation can arise both from a sextuplet (triplet) with
or without the tetrahedron property since the tetrahedron property
\emph{is a property of the sextuplet $($triplet$)$ of equations and not a property
of the equation itself}.
Examples of this phenomenon were already known in literature
\cite{ABS2009,Atkinson2008, HietarintaViallet2012}.
The example presented here is another one and it is rather non-trivial.

We stress out that since the non-autonomous Hietarinta equation~\eqref{eq:j1na}
is linearizable its Lax pairs obtained both
from triplet~\eqref{eq:j1nasystem} and from
reversing the transformation~\eqref{eq:mobnm} in the triplet~\eqref{eq:j2vsystem2} should both be fake.
Being fake these two Lax pair cannot give much information.
However also integrable equations can have, in principle,
Lax pairs coming from sextuplet (triplet) without tetrahedron property.
Given a quad-equation with such properties
it will be an interesting issue to investigate
the relationship between these Lax pairs as it was
done for the continuous Painlev\'e~I and~II equations~\cite{Joshietal2009}.
We remark that in this case we are ensured that there
is no point transformation between the sextuplet \eqref{eq:j1hietsystem}
and the triplet \eqref{eq:j1hietsystemm} since the f\/irst one does
not possess the tetrahedron property while the second one does.
On the other hand it is possible to prove the triplet
\eqref{eq:j1nasystem} and the triplet \eqref{eq:j2vsystem}
are not related by a $\Mob^{8}$ transformation even though
they both do not possess the tetrahedron property.

\begin{Remark}
 We observe also that the M\"obius transformation
 \begin{gather}
 y_{n,m} = \frac{u_{n,m}+K_{n,m}^{(1)}}{u_{n,m}+K_{n,m}^{(2)}} \label{eq:moby}
 \end{gather}
 bring equation \eqref{eq:j1na} into another autonomous quad-equation
 \begin{gather*}
 y_{n,m}y_{n+1,m}y_{n,m+1}y_{n+1,m+1}=1. 
 \end{gather*}
 This other equation is linked with equation \eqref{eq:j2v}
 through the non-autonomous M\"obius transformation~\cite{Hietarinta2005}
 \begin{gather*}
 y_{n, m} = \big( \Fppp+\Fmmp \big)v_{n, m}+\frac{\Fmpm+\Fpmm}{v_{n, m}}. 
 \end{gather*}
\end{Remark}

The existence of the point transformations \eqref{eq:mobnm}
and \eqref{eq:moby} is in itself suf\/f\/icient to explain the linearizability
of the non-autonomous Hietarinta equation \eqref{eq:j1na},
since equation \eqref{eq:j2v} is linearizable,
Darboux integrable and its generalized symmetries of every order
are known \cite{GSL_Pavel}.
These properties can then be inferred through this point
transformation, but for the rest of this paper we have chosen
to present a direct derivation.
Our choice is motivated from the fact that equation \eqref{eq:j1na}
can be seen as a simple, yet nontrivial, example of all the properties
found in the study of the trapezoidal $H^{4}$ and $H^{6}$ equations.

\section{Darboux integrability}\label{sec:darb}

Suppose we are given a quad-equation, possibly non-autonomous
\begin{gather}\label{eq:quadequana}
 Q_{n,m}( u_{n,m},u_{n+1,m},u_{n,m+1},u_{n+1,m+1} )=0.
\end{gather}
We say that such a quad-equation is \emph{Darboux integrable}
if there exist two independent \emph{first integrals},
one containing only shifts in $n$, and the other containing only shifts in~$m$,
i.e., that there exist two independent functions
 \begin{gather*}
 W_1=W_{1,n,m}(u_{n+l_1,m},u_{n+l_1+1,m},\ldots,u_{n+k_1,m}), \label{eq:darbfint1} \\
 W_2=W_{2,n,m}(u_{n,m+l_2},u_{n,m+l_2+1},\ldots,u_{n,m+k_2}), \label{eq:darbfint2}
 \end{gather*}
where $l_1<k_1$ and $l_2<k_2$ are integers, such that the relations
\begin{subequations}\label{eq:darbdef}
 \begin{gather*}
 (T_n-\Id)W_2=0, \label{eq:darb1} \\
 (T_m-\Id)W_1=0,  \label{eq:darb2}
 \end{gather*}
\end{subequations}
where $T_n h_{n,m}=h_{n+1,m}$, $T_m h_{n,m}=h_{n,m+1}$,
and $\Id h_{n,m}=h_{n,m}$, hold true identically on the solutions of~\eqref{eq:quadequana}.
The number $k_{i}-l_{i}$, where $i=1,2$, is called the \emph{order} of
the f\/irst integral $W_{i}$.

Any Darboux integrable equation is linearizable~\cite{AdlerStartsev1999}.
Indeed let us introduce two new f\/ields $\tilde{u}_{n,m}=W_{1,n,m}$ and $\hat{u}_{n,m}=W_{2,n,m}$. Then
\begin{subequations} \label{eq:fintlin}
 \begin{gather}
 u_{n,m} \to \tilde u_{n,m}, \label{eq:fintlinA} \\
 u_{n,m} \to \hat u_{n,m}, \label{eq:fintlinB}
 \end{gather}
\end{subequations}
def\/ine two non-point transformations of the f\/ield $u_{n,m}$ into $\tilde{u}_{n,m}$ and $\hat{u}_{n,m}$.
Moreover these two new f\/ields satisfy two \emph{trivial linear equations}
\begin{subequations} \label{eq:fintlin2}
 \begin{gather}
 \tilde u_{n,m+1} - \tilde u_{n,m} = 0 , \label{eq:fintlin2A} \\
 \hat u_{n+1,m} - \hat u_{n,m} = 0. \label{eq:fintlin2B}
 \end{gather}
\end{subequations}
Therefore we can conclude that any Darboux integrable equation is \emph{linearizable in two different ways}, i.e.,
using transformation \eqref{eq:fintlinA} bringing to \eqref{eq:fintlin2A} or using the transformation~\eqref{eq:fintlinB} bringing to~\eqref{eq:fintlin2B}.

Methods for calculating f\/irst integrals of non-autonomous quad-equations~\eqref{eq:quadequana} with two-periodic coef\/f\/icients were given in \cite{GarifullinYamilov2015,GSY_DarbouxI}. In particular in \cite{GSY_DarbouxI} was presented a~new algorithm that relies on the fact that in the case of non-autonomous quad-equations \eqref{eq:quadequana} with two-periodic coef\/f\/icients we can, in general, represent the f\/irst integrals in the form{\samepage
\begin{gather*}
 W_{i} = \Fppp W_{i}^{(+,+)}+\Fmpm W_{i}^{(-,+)}
 +\Fpmm W_{i}^{(+,-)} + \Fmmp W_{i}^{(-,-)}, 
\end{gather*}
where $F_{k}^{(\pm)}$ are given by \eqref{eq:fk} and the $W_{i}^{(\pm,\pm)}$ are functions.}

Applying the algorithm presented in \cite{GSY_DarbouxI}
we f\/ind that the non-autonomous Hietarinta equation
\eqref{eq:j1na} possesses two f\/irst order f\/irst integrals
in both directions
\begin{subequations} \label{eq:j1nafint}
 \begin{gather*}
 W_{1} = \alpha_{1} W_{1}^{(\alpha_{1})}+\beta_{1} W_{1}^{(\beta_{1})}, \label{eq:j1naW1} \\
 W_{2} = \alpha_{2} W_{2}^{(\alpha_{2})}+\beta_{2} W_{2}^{(\beta_{2})}, \label{eq:j1naW2}
 \end{gather*}
\end{subequations}
where
 \begin{gather*}
 W_{1}^{(\alpha_{1})} =F_n^{(+)}F_m^{(+)}{\frac { ( u_{{n,m}}+e_2 )
 ( u_{{n+1,m}}+o_1) }{
 ( u_{{n,m}}+ e_1 )
 ( u_{{n+1,m}}+e_2 ) }}
 +F_n^{(+)}F_m^{(-)} {\frac { ( u_{{n,m}}+o_2 )
 ( u_{{n+1,m}}+o_1 ) }{%
 ( u_{{n,m}}+e_1 )
 ( u_{{n+1,m}} +o_{2}) }}, \label{eq:j1naW1a} \\
 W_{1}^{(\beta_{1})}=F_n^{(-)}F_m^{(+)} {\frac { ( u_{{n,m}}+o_1 )
 ( u_{{n+1,m}}+ e_2 ) }{
 ( u_{{n,m}}+e_2 )
 ( u_{{n+1,m}}+e_1 ) }}+F_n^{(-)}F_m^{(-)} {\frac { ( u_{{n,m}}+o_1 )
 ( o_2+u_{{n+1,m}} ) }{%
 ( u_{{n,m}}+ o_2 ) ( u_{{n+1,m}}+e_1 ) }}, \label{eq:j1naW1b} \\
 W_{2}^{(\alpha_{2})}
 =F_n^{(+)}F_m^{(+)} {\frac { ( u_{{n,m}}+e_2 )
 ( u_{{n,m+1}}+e_1 ) }{%
 ( u_{{n,m}}+e_1 )
 ( u_{{n,m+1}}+o_2 ) }}+F_n^{(-)}F_m^{(+)} {\frac { ( u_{{n,m}}+e_2 )
 ( o_1+u_{{n,m+1}} ) }{
 ( u_{{n,m}}+o_1 )
 ( u_{{n,m+1}}+o_2 ) }}, \label{eq:j1naW2a} \\
 W_{2}^{(\beta_{2})}
 =F_n^{(+)}F_m^{(-)} \frac { ( u_{{n,m}}+o_2 )
 ( u_{{n,m+1}}+e_1 ) }{
 ( u_{{n,m}}+e_1 )
 ( u_{{n,m+1}}+e_2 )}
 +F_n^{(-)}F_m^{(-)} {\frac { ( u_{{n,m}}+o_2 )
 ( u_{{n,m+1}}+o_{1} ) }{%
 ( u_{{n,m}}+o_{1} )
 ( u_{{n,m+1}}+ e_2 ) }}.
 \label{eq:j1naW2b}
 \end{gather*}

\section{Generalized symmetries}\label{sec:gensymm}

A vector f\/ield of the form
\begin{gather}
 \hat X = g_{n,m}\big( \mathbf{u}_{n,m}^{D} \big) \partial_{u_{n,m}},
 \qquad \mathbf{u}_{n,m}^{D}
 = \{ u_{n+i,m+j} \}_{i=l_{1},\dots,k_{1},j=l_{2},\dots,k_{2}},\label{eq:infgenquad}
\end{gather}
where $l_{1}<k_{1}$ and $l_{2}<k_{2}$ is said to be a \emph{generalized symmetry} for the quad-equation \eqref{eq:quadequana} if its discrete
prolongation
\begin{gather*}
\mathrm{pr}^{D}\hat X = g_{n,m}\partial_{u_{n,m}} +T_{n}g_{n,m}\partial_{u_{n+1,m}}+T_{m}g_{n,m}\partial_{u_{n,m+1}}+T_{n}T_{m}g_{n,m}\partial_{u_{n+1,m+1}},
\end{gather*}
is such that
\begin{gather*}
g_{n,m} \frac{\partial Q_{n,m}}{\partial u_{n,m}}
 +T_n g_{n,m} \frac{\partial Q_{n,m}}{\partial u_{n+1,m}}
 + T_m g_{n,m} \frac{\partial Q_{n,m}}{\partial u_{n,m+1}}
 + T_n T_m g_{n,m}\frac{\partial Q_{n,m}}{\partial u_{n+1,m+1}} =0,
\end{gather*}
identically on the solutions of \eqref{eq:quadequana} \cite{Garifullin2012}. Symmetries of this kind with $k_{i}=-l_{i}=1$, i.e., three-point generalized symmetries were f\/irst considered in \cite{Hydon2007}. Moreover three-point generalized symmetries were threated in \cite{LeviYamilov2009,LeviYamilov2011}, and the general case was discussed in \cite{Garifullin2012}. In \cite{Garifullin2012} it was also proved that quad-equations \eqref{eq:quadequana} do not posses ``mixed'' symmetries, i.e., the function~$g_{n,m}$ in~\eqref{eq:infgenquad}, known as the \emph{characteristic of the generalized symmetry}, is the sum of two simpler functions depending only on variables shifted only in one direction
\begin{gather*}
 g_{n,m}\big( \mathbf{u}_{n,m}^{D} \big) = g_{n,m}^{(1)} ( u_{n+l_{i},m},\dots,u_{n+k_{1},m} ) + g_{n,m}^{(2)}( u_{n,m+l_{2}},\dots,u_{n,m+k_{2}}). 
\end{gather*}

The relationship between generalized symmetries and Darboux integrability is well known in the continuous case~\cite{ZhiberSokolov2011} and in the discrete case have been investigated in \cite{AdlerStartsev1999,Startsev2014,Startsev2016}. As conjectured in~\cite{GSL_Pavel} and proved in~\cite{Startsev2016},
a quad-equation is Darboux integrable \emph{if and only if}
it possesses generalized symmetries depending
on arbitrary function in the following form
\begin{subequations} \label{eq:gsymm}
 \begin{gather}
 g^{(1)}_{n,m} = R^{(1)}\big(F_{n}\big( T^{p_{1}}_{n}W_{1},\dots,T^{q_{1}}_{n}W_{1}\big)\big), \\
 g^{(2)}_{n,m} = R^{(2)} \big(G_{m}\big( T^{p_{2}}_{m}W_{2},\dots,T^{q_{2}}_{m}W_{2}\big)\big),
 \end{gather}
\end{subequations}
where $p_{i}<q_{i}$ and $R^{(i)}$ are operators of the form
\begin{subequations} \label{eq:Ri}
 \begin{gather}
 R^{(1)} = \sum_{r=j_{1}}^{h_{1}} \lambda_{r} ( u_{n+l_{i}',m},\dots,u_{n+k_{1}',m} )T_{n}^{r}, \label{eq:R1} \\
 R^{(2)} = \sum_{r=j_{2}}^{h_{2}} \mu_{r} ( u_{n,m+l_{2}'},\dots,u_{n,m+k_{2}'} )T_{m}^{r}, \label{eq:R2}
 \end{gather}
\end{subequations}
with $j_{i}<h_{i}$ and $l_{i}'<k_{i}'$. The operators $R^{(i)}$ \eqref{eq:Ri} are called \emph{symmetry drivers}.

It is then easy to show that the non-autonomous Hietarinta equation possesses the following two-point generalized symmetries, i.e., such that $k_{i}=1$, $l_{i}=0$ in \eqref{eq:infgenquad}
 \begin{gather*}
 g^{(1)}_{n,m} =
 \left[
 \Fppp{\frac { ( u_{{n,m}}+{ e_1} )
 ( u_{{n,m}}+{ e_2}) }{ { e_1}-{ e_2} }}
 +\Fpmm
 {\frac { ( u_{{n,m}}+{ e_1} ) ( u_{{n,m}}+{ o_2} )}{ { e_1}-{ o_2} }}
 \right]\!
 {F}^{(1)}_{{n}} \big( W_{1}^{(\alpha_{1})} \big)\nonumber \\
 \quad{}+\left[\Fmpm{\frac { ( u_{{n,m}} + { e_2} ) ( u_{{n,m}} + {o_1} )}{
 { e_2}-{ o_1} }}
 - \Fmmp{\frac { ( u_{{n,m}} + { o_2} )
 ( { o_1} + u_{{n,m}} ) }{{o_1} -{o_2}}}\right]
 { G}_{{n}}^{(1)} \big( W_{1}^{(\beta_{1})} \big),      \label{j1nagsn} \\
 g^{(2)}_{n,m} =\left[
 {\Fppp\frac { ( u_{{n,m}}+{ e_1} ) ( u_{{n,m}}+{ e_2} )}{
 { e_1}-{ e_2} }}
 -\Fmpm{\frac { ( u_{{n,m}}+{ e_2} )
 ( u_{{n,m}} +{ o_1}) }{
 { e_2}-{ o_1} }}\right]\!
 { F}_{{m}}^{(2)} \big( W_{2}^{(\alpha_{1})} \big)\nonumber \\
  \quad{}+\left[\Fpmm{\frac { ( u_{{n,m}} + { e_1})
 ( u_{{n,m}} + { o_2} )}{ ( { e_1}-{ o_2} )}}
 + \Fmmp{\frac { ( u_{{n,m}} + { o_2} )
 ( u_{{n,m}} + { o_1} )}{ { o_1} -{ o_2}}}\right]
 { G}_{{m}}^{(2)} \big( W_{2}^{(\alpha_{1})} \big),      \label{j1nagsm}
 \end{gather*}
where $F^{(i)}_{k}$ and $G_{k}^{(i)}$ are arbitrary functions of their arguments. This implies that we have two multipliticative symmetry drivers $R^{(i)}$ \eqref{eq:Ri} in each direction
 \begin{gather*}
 R^{(1,\alpha_{1})} = \Fppp{\frac { ( u_{{n,m}}+{ e_1} ) ( u_{{n,m}}+{ e_2}) }{ { e_1}-{ e_2} }} +\Fpmm
 {\frac { ( u_{{n,m}}+{ e_1} ) ( u_{{n,m}}+{ o_2} )}{ { e_1}-{ o_2} }}, \label{j1naR1a} \\
 R^{(1,\beta_{1})}= \Fmpm{\frac { ( u_{{n,m}}+{ e_2} ) ( u_{{n,m}}+{o_1} )}{ { e_2}-{ o_1} }} -\Fmmp{\frac { ( u_{{n,m}}+{ o_2} ) ( { o_1}+u_{{n,m}} ) }{{o_1} -{o_2}}}, \label{j1naR1b} \\
 R^{(2,\alpha_{2})} = {\Fppp\frac { ( u_{{n,m}}+{ e_1} ) ( u_{{n,m}}+{ e_2} )}{ { e_1}-{ e_2} }} -\Fmpm{\frac { ( u_{{n,m}}+{ e_2} ) ( u_{{n,m}} +{ o_1}) }{ { e_2}-{ o_1} }}, \label{j1naR2a} \\
 R^{(2,\beta_{2})} =\Fpmm{\frac { ( u_{{n,m}}+{ e_1}) ( u_{{n,m}}+{ o_2} )}{ ( { e_1}-{ o_2} )}} +\Fmmp{\frac { ( u_{{n,m}}+{ o_2} )
 ( u_{{n,m}}+{ o_1} )}{ { o_1} -{ o_2}}}, \label{j1naR2b}
 \end{gather*}
and then that the generalized symmetry of arbitraryorder $N_{i}=q_{i}-p_{i}$ are given by \eqref{eq:gsymm}
 \begin{gather*}
 g^{(1)}_{n,m} =
 R^{(1,\alpha_{1})}
 {F}^{(1)}_{{n}} \big(T_{n}^{p_{1}}W_{1}^{(\alpha_{1})},\dots,T_{n}^{q_{1}} W_{1}^{(\alpha_{1})} \big)
+
R^{(1,\beta_{1})} {G}^{(1)}_{{n}} \big(T_{n}^{p_{1}}W_{1}^{(\beta_{1})},\dots,T_{n}^{q_{1}} W_{1}^{(\beta_{1})} \big), \\ 
 g^{(2)}_{n,m} =
 R^{(2,\alpha_{2})} {F}^{(2)}_{{n}} \big(T_{n}^{p_{2}}W_{2}^{(\alpha_{2})},\dots,T_{n}^{q_{2}} W_{2}^{(\alpha_{2})} \big)
+ R^{(2,\beta_{2})}
 {G}^{(2)}_{{n}} \big(T_{n}^{p_{2}}W_{2}^{(\beta_{2})},\dots,T_{n}^{q_{2}} W_{2}^{(\beta_{2})} \big), 
 \end{gather*}
where $F_{k}^{(i)}( x_{1},\dots,x_{N_{i}} )$ and $G_{k}^{(i)}( x_{1},\dots,x_{N_{i}})$ are arbitrary functions of their arguments.

\section{General solutions} \label{sec:sol}

In this section we construct the general solution of the non-autonomous Hietarinta equa\-tion \eqref{eq:j1na} using the method presented in~\cite{GSY_DarbouxII, GSY_DarbouxI}, which is a modif\/ication of the procedure presented in~\cite{GarifullinYamilov2012}.
By general solution we mean a representation of the solution of a quad-equation~\eqref{eq:quadequana} in terms of the right number of arbitrary functions of one lattice variable~$n$ or~$m$.
Since quad-equations are the discrete analogue of
second order hyperbolic partial dif\/ferential equations,
the general solution must contain an arbitrary function
in the~$n$ direction and another one in the $m$ direction.

To obtain the solution we will need only the $W_{1}$ integrals we
derived in Section \ref{sec:darb}
and the fact that the relation \eqref{eq:darb2}
implies $W_{1}=\xi_{n}$ with $\xi_{n}$ an arbitrary function of $n$.
The equation $W_{1}=\xi_{n}$ is an
ordinary dif\/ference equation in the $n$ direction depending
parametrically on $m$.
Then from every $W_{1}$ integral we can derive
two dif\/ferent ordinary dif\/ference equations, one
corresponding to $m$ even and one corresponding to $m$ odd.
In both the resulting equations we can get rid of the two-periodic
terms by considering the cases $n$ even and $n$ odd and using the
def\/initions
\begin{subequations} \label{eq:genautsub}
 \begin{gather}
 u_{2k,2l} = v_{k,l}, \qquad u_{2k+1,2l} = w_{k,l}, \label{eq:genautsubl} \\
 u_{2k,2l+1} = y_{k,l}, \qquad u_{2k+1,2l+1} = z_{k,l}. \label{eq:genautsublp}
 \end{gather}
\end{subequations}
This transformation brings both equations into a \emph{system} of \emph{coupled difference equations}.
Apply the even/odd splitting \eqref{eq:genautsub} of the lattice variables to describe a general solution we will need two arbitrary functions in both directions, i.e., we will need a total of four arbitrary functions.

We assume without loss of generality $\alpha_{1}=\beta_{1}=1$ in \eqref{eq:j1naW1}
\begin{gather}
F_n^{(+)}F_m^{(+)}{\frac { ( u_{{n,m}}+e_2 )
 ( u_{{n+1,m}}+o_1) }{ ( u_{{n,m}}+ e_1 )
 ( u_{{n+1,m}}+e_2 ) }}
+F_n^{(+)}F_m^{(-)} {\frac { ( u_{{n,m}}+o_2 )
 ( u_{{n+1,m}}+o_1 ) }{%
 ( u_{{n,m}}+e_1 )
 ( u_{{n+1,m}} +o_{2}) }} \nonumber\\
 \quad {}+F_n^{(-)}F_m^{(+)} {\frac { ( u_{{n,m}}+o_1 )
 ( u_{{n+1,m}}+ e_2 ) }{%
 ( u_{{n,m}}+e_2 )
 ( u_{{n+1,m}}+e_1 ) }}
 +F_n^{(-)}F_m^{(-)} {\frac { ( u_{{n,m}}+o_1 )
 ( o_2+u_{{n+1,m}} ) }{%
 ( u_{{n,m}}+ o_2 ) ( u_{{n+1,m}}+e_1 ) }}=\xi_{n}.
 \label{eq:j1naW1eq}
\end{gather}
Then we treat the cases $m$ even and odd separately.

Case $m=2l$: In this case equation \eqref{eq:j1naW1eq} becomes
 \begin{gather}
F_n^{(+)}{\frac { ( u_{{n,2l}}+e_2 ) ( u_{{n+1,2l}}+o_1) }{ ( u_{{n,2l}}+ e_1 )
 ( u_{{n+1,2l}}+e_2 ) }}+F_n^{(-)} {\frac { ( u_{{n,2l}}+o_1 ) ( u_{{n+1,2l}}+ e_2 ) }{ ( u_{{n,2l}}+e_2 )
 ( u_{{n+1,2l}}+e_1 ) }} =\xi_{n}.
 \label{eq:j1eQm}
 \end{gather}
 We can then separate the even and odd part in $n$ of equation \eqref{eq:j1eQm} and apply the transforma\-tion~\eqref{eq:genautsubl}.
 We obtain the following system
 \begin{subequations} \label{eq:j1naeQmsys}
 \begin{gather}
 {\frac { ( v_{{k,l}}+{ e_2} )
 ( w_{{k,l}}+{ o_1} ) }{ ( v_{{k,l}}+{ e_1} )
 ( w_{{k,l}}+{ e_2} ) }} =\xi_{{2 k}}, \label{eq:j1naeQmk2} \\
 {\frac { ( w_{{k,l}}+{ o_1} )
 ( v_{{k+1,l}}+{ e_2} ) }{%
 ( w_{{k,l}}+{ e_2} )
 ( v_{{k+1,l}}+{ e_1} ) }} =\xi_{{2 k+1}}. \label{eq:j1naeQmkp2}
 \end{gather}
 \end{subequations}
 This system is linear and equation \eqref{eq:j1naeQmk2} is not a dif\/ference equation, but def\/ines
 $w_{k,l}$ in terms of~$v_{k,l}$
 \begin{gather}
 w_{{k,l}}=-{\frac { ( -{ o_1}+\xi_{{2 k}}{ e_2} ) v_{{k,l}}
 -{ e_2} { o_1}+\xi_{{2 k}}{ e_1} { e_2}}{%
( -1+\xi_{{2 k}} ) v_{{k,l}}-{ e_2}+\xi_{{2 k}}{ e_1}}}. \label{eq:j1nawkldef}
 \end{gather}
 Inserting $w_{k,l}$ from \eqref{eq:j1nawkldef} into
 equation \eqref{eq:j1naeQmkp2} we obtain that
 $v_{k,l}$ solves the following discrete Riccati equation
 \begin{gather}
 v_{{k+1,l}}={\frac {
( -\xi_{{2 k}}{ e_2}+\xi_{{2 k+1}}{ e_1} )
 v_{{k,l}}
 -\xi_{{2 k}}{ e_1} { e_2}+\xi_{{2 k+1}}{ e_1} { e_2}}{
( \xi_{{2 k}}-\xi_{{2 k+1}} )
 v_{{k,l}}-\xi_{{2 k+1}}{ e_2}+\xi_{{2 k}}{ e_1}}}. \label{eq:j1navkleq}
 \end{gather}
 Equation \eqref{eq:j1navkleq} is linearized through the M\"obius transformation
 \begin{gather}
 v_{{k,l}}=-{ e_2}-{\frac {{ e_2}}{V_{{k,l}}}}, \label{eq:j1naVkldef}
 \end{gather}
 and yields the following linear equation for $V_{k,l}$
 \begin{gather}
 V_{{k+1,l}}={\frac {\xi_{{2 k}}}{\xi_{{2 k+1}}}}V_{{k,l}}
 -{\frac {{ e_2} ( \xi_{{2 k}}-\xi_{{2 k+1}} ) }{
 \xi_{{2 k+1}} ( { e_1}-{ e_2} ) }}. \label{eq:j1naVkleq}
 \end{gather}
 Exploiting the arbitrariness of $\xi_{2k}$ we introduce a new arbitrary function $a_{k}$ by def\/ining
 \begin{gather}
 \xi_{{2 k}}={\frac {a_{{k}}}{a_{{k+1}}}}\xi_{{2 k+1}}. \label{eq:j1nalkdef}
 \end{gather}
 Then equation \eqref{eq:j1naVkleq} becomes
 \begin{gather}
 a_{{k+1}}V_{{k+1,l}} =a_{{k}}V_{{k,l}} -{\frac {{ e_2} ( a_{{k}}-a_{{k+1}} ) }{{ e_1} -{ e_2}}}. \label{eq:j1naVkleq2}
 \end{gather}
 Equation \eqref{eq:j1naVkleq2} is a \emph{total difference} hence its solution is
 \begin{gather}
 V_{{k,l}}={\frac {\alpha_{{l}}}{a_{{k}}}} +{\frac {{ e_2}}{{ e_1}-{ e_2}}}. \label{eq:j1naVklsol}
 \end{gather}
 Then inserting \eqref{eq:j1naVklsol} into \eqref{eq:j1naVkldef} and \eqref{eq:j1nawkldef} we obtain the solution of the system \eqref{eq:j1naeQmsys}
 \begin{subequations} \label{eq:j1naeQmsol}
 \begin{gather}
 v_{{k,l}} =-{\frac {{ e_2} ( \alpha_{{l}}
 ( { e_1}-{ e_2} ) +a_{{k}}{ e_1} ) }{
 \alpha_{{l}} ( { e_1}-{ e_2} ) +a_{{k}}{ e_2}}}, \label{eq:j1nawklsol} \\
 w_{{k,l}}=-{\frac {{ e_2} ( \xi_{{2 k+1}}\alpha_{{l}}{ e_1}+{ o_1} a_{{k+1}}
 -\xi_{{2 k+1}}{ e_2} \alpha_{{l}} ) }{ ( a_{{k+1}}-\xi_{{2 k+1}}\alpha_{{l}} )
 { e_2}+\xi_{{2 k+1}} \alpha_{{l}}{ e_1}}}. \label{eq:j1nawklsol2}
 \end{gather}
 \end{subequations}

Case $m=2l+1$: In this case equation \eqref{eq:j1naW1eq} becomes:
 \begin{gather}
F_n^{(+)} {\frac { ( u_{{n,2l+1}}+o_2 )
 ( u_{{n+1,2l+1}}+o_1 ) }{ ( u_{{n,2l+1}}+e_1 )
 ( u_{{n+1,2l+1}} +o_{2}) }}
 +F_n^{(-)} {\frac { ( u_{{n,2l+1}}+o_1 )
 ( o_2+u_{{n+1,2l+1}} ) }{%
 ( u_{{n,2l+1}}+ o_2 ) ( u_{{n+1,2l+1}}+e_1 ) }}=\xi_{n}. \label{eq:j1naePm}
 \end{gather}
 We can then separate the even and odd part in
 $n$ of equation \eqref{eq:j1naePm} and apply the
 transformation~\eqref{eq:genautsublp}.
 We obtain the following system
 \begin{subequations} \label{eq:j1naePmsys}
 \begin{gather}
 {\frac { ( y_{{k,l}}+{ o_2} )
 ( z_{{k,l}}+{ o_1} ) }{
 ( y_{{k,l}}+{ e_1} )
 ( z_{{k,l}}+{ o_2} ) }}
 ={\frac {a_{{k}}}{a_{{k+1}}}}\xi_{{2 k+1}}, \label{eq:j1naePmk2}
 \\
 {\frac { ( z_{{k,l}}+{ o_1} )
 ( y_{{k+1,l}}+{ o_2} ) }{%
 ( z_{{k,l}}+{ o_2} )
 ( y_{{k+1,l}}+{ e_1} ) }} =\xi_{{2 k+1}},
 \label{eq:j1naePmkp2}
 \end{gather}
 \end{subequations}
where we used the def\/inition of $\xi_{2k}$ \eqref{eq:j1nalkdef}. This system is linear and equation \eqref{eq:j1naePmk2}
 is not a~dif\/ference equation, but def\/ines $z_{k,l}$ in terms of $y_{k,l}$
 \begin{gather}
 z_{{k,l}}={\frac {-a_{{k}}{ o_2} ( y_{{k,l}}+{ e_1} )
 \xi_{{2 k+1}}+{ o_1} a_{{k+1}} ( y_{{k,l}}+{ o_2} ) }{%
 a_{{k}} ( y_{{k,l}}+{ e_1} ) \xi_{{2 k+1}}
 -a_{{k+1}} ( y_{{k,l}}+{ o_2} ) }}.
 \label{eq:j1nazklsol}
 \end{gather}
 Inserting $z_{k,l}$ from \eqref{eq:j1nazklsol} into
 equation \eqref{eq:j1naePmkp2} we obtain that
 $y_{k,l}$ solves the following discrete Riccati
 equation
 \begin{gather}
 y_{{k+1,l}}=
 {\frac {{ e_1} ( y_{{k,l}}+ { o_2}) a_{{k+1}}
 -a_{{k}}{ o_2} ( y_{{k,l}}+{ e_1} ) }{%
 -( y_{{k,l}}+{ o_2} ) a_{{k+1}}
 +a_{{k}} ( y_{{k,l}}+{ e_1} ) }}.
 \label{eq:j1naykleq}
 \end{gather}
 Equation \eqref{eq:j1naykleq} is linearized through the M\"obius transformation
 \begin{gather}
 y_{{k,l}}=-{ o_2}-{\frac {{ o_2}}{Y_{{k,l}}}}, \label{eq:j1naYkldef}
 \end{gather}
 which yields the following equation for $Y_{k,l}$
 \begin{gather}
 a_{{k+1}}Y_{{k+1,l}}=a_{{k}}Y_{{k,l}} -{\frac {{ o_2} ( a_{{k}}-a_{{k+1}} ) }{{ e_1}}-{ o_2}}. \label{eq:j1naYkleq}
 \end{gather}
 Equation \eqref{eq:j1naYkleq} is already a~total dif\/ference, so its solution is
 \begin{gather}
 Y_{{k,l}}={\frac {\beta_{{l}}}{a_{{k}}}} +{\frac {{ o_2}}{{ e_1}-{ o_2}}} . \label{eq:j1naYklsol}
 \end{gather}
 Then inserting \eqref{eq:j1naYklsol} into \eqref{eq:j1naYkldef} and \eqref{eq:j1nazklsol} we obtain the solution of the system \eqref{eq:j1naePmsys}
 \begin{subequations} \label{eq:j1naePmsol}
 \begin{gather}
 y_{{k,l}}= -{\frac {{ o_2} ( \beta_{{l}} ( { e_1}-{ o_2} )
 +a_{{k}}{ e_1} ) }{ \beta_{{l}} ( { e_1}-{ o_2} )
 +a_{{k}}{ o_2}}}, \label{eq:j1nayklsol} \\
 z_{{k,l}}=-{\frac {{ o_2} ( \xi_{{2 k+1}}\beta_{{l}}{ e_1}
 -\beta_{{l}}\xi_{{2 k+1}}{ o_2}+{ o_1} a_{{k+1}} ) }{%
 ( a_{{k+1}}-\beta_{{l}}\xi_{{2 k+1}} ) { o_2}
 +\xi_{{2 k+1}}\beta_{{l}}{ e_1}}}. \label{eq:j1nazklsol2}
 \end{gather}
 \end{subequations}

The solution of the non-autonomous Hietarinta equation \eqref{eq:j1na} is then given by formulas~\eqref{eq:j1naeQmsol} and~\eqref{eq:j1naePmsol}. The four arbitrary functions are $\xi_{2k+1}$, $a_{k}$, $\alpha_{l}$ and $\beta_{l}$. It can be checked that this is the general solution by inserting it into \eqref{eq:j1na}

\section{Conclusions}\label{sec:conclusions}

In this paper we presented a new approach to the Hietarinta equation \eqref{eq:j1}. From the consi\-de\-ration that the Hietarinta equation possesses the property of the consistency around the cube~\cite{Hietarinta2005}, but does not possess the discrete symmetries of the square~\eqref{eq:squaresymm} we applied to it the Boll's construction developed in~\cite{ABS2009,Boll2011,Boll2012a,Boll2012b,HietarintaViallet2012,Xenitidis2009}. The result was a seemingly new non-autonomous, two-periodic quad-equation which we called the non-autonomous Hietarinta equation~\eqref{eq:j1na}. In this sense we ``reconstructed'' the Hietarinta equation~\eqref{eq:j1}, since from the same single-cell equation instead of using the standard embedding~\eqref{eq:trivialemb} we adopted a~dif\/ferent one resulting in a dif\/ferent equation on the $\Z^{2}$ lattice.

We devoted the rest of the paper to the study of the integrability properties of this equation.
The results of Sections~\ref{sec:entropy}--\ref{sec:sol}
can be seen as a nice example of the properties that were
found for the two classes of linearizable equations belonging
to Boll's classif\/ications, namely the trapezoidal~$H^{4}$
equations and the $H^{6}$ equations \cite{GSL_general,GSL_Gallipoli15,GSL_Pavel,GSL_symmetries,GSL_QV,GSY_DarbouxII,
GSY_DarbouxI}.
Indeed we proved in Section~\ref{sec:qvmobius} that the
non-autonomous Hietarinta equation can be mapped into a
particular case of the~$_{1}D_{4}$ equation, an equation
belonging to the~$H^{6}$ class.
We remark that the identif\/ication of the non-autonomous
Hietarinta equation~\eqref{eq:j1na} with equation~\eqref{eq:j2v}
is possible \emph{only} in the extended framework of
the consistency around the cube given in~\cite{ABS2009,Boll2011,Boll2012a,Boll2012b},
since in order to transform the non-autonomous
Hietarinta equation~\eqref{eq:j1na} into~\eqref{eq:j2v}
the non-autonomous M\"obius transformation~\eqref{eq:mobnm}
must be used.
In the framework of \cite{ABS2003} this is not possible
since transformations of this kind are not allowed.
Therefore the construction carried out in this paper
is ``necessary'' to obtain this result.

In this work we prove that the problem of the embedding
of single cell quad-equations possessing the
consistency around the cube is crucial.
As already discussed in \cite{HietarintaViallet2012} there might
exist multiple embeddings preserving on the whole
lattice the consistency around the cube.
To understand how many and what they are is then a relevant task.
We conjecture that in the procedure of extension might
be essential to form \emph{bi-quadratic patterns}
\cite{Boll2011,Boll2012a,Boll2012b}.
To form a~bi-quadratic pattern means that, given a quad-equation
on the single cell~\eqref{eq:quadequa}, to construct an equation
on the~$\Z^{2}$ lattice in a way such that
the bi-quadratics~\eqref{eq:biquadr} pertaining to the sides
of neighboring cells are of the same type.
However preserving the bi-quadratic patterns might be inef\/fectual
in the case of the~$H^{6}$ due to the lack of a well-def\/ined
``black/white'' assignment to the vertexes of the elementary cell.

In Section \ref{sec:qvmobius} we discussed, exploiting the
M\"obius transformation \eqref{eq:mobnm}, how the non-auto\=no\-mous
Hietarinta equation \eqref{eq:j1na} can arise as bottom equation
from a triplet \emph{possessing} the tetrahedron property
or from a triplet \emph{without} tetrahedron property.
This leaves open the problem if any other known quad-equation
may arise from triplet without tetrahedron property
and what kind of information such non-tetrahedron triplet can give.

Finally we point out that whereas it was very easy
to prove the Darboux integrability for the non-autonomous
Hietarinta equation \eqref{eq:j1na}, if an analogous
result holds for the Hietarinta equation \eqref{eq:j1}
with the standard embedding \eqref{eq:trivialemb} it is not known.
It was conjectured in \cite{GubbiottiPhD2017} that
the Hietarinta equation \eqref{eq:j1} with the standard embedding
\eqref{eq:trivialemb} is Darboux integrable with
f\/irst integrals of order greater than two,
but technical dif\/f\/iculties prevent a full proof
of this statement.

\subsection*{Acknowledgements}

We would like to thank Professor Decio Levi for the many interesting and fruitful discussion during the preparation of this paper. We thank the anonymous referees for their suggestions on how to improve the paper.
GG is supported by INFN IS-CSN4 \emph{Mathematical Methods of Nonlinear Physics} and by the Australian Research Council through an Australian Laureate Fellowship grant FL120100094.

\pdfbookmark[1]{References}{ref}
\LastPageEnding

\end{document}